\documentclass[aps,prd,twocolumn,amssymb,floatfix,superscriptaddress,nofootinbib,longbibliography,noeprint]{revtex4-1}

\usepackage{graphicx}
\usepackage{dcolumn}
\usepackage{bm}
\usepackage{latexsym}
\usepackage{color}
\usepackage{cancel}
\usepackage[colorlinks=true,linkcolor=blue,citecolor=blue,urlcolor=magenta]{hyperref}
\usepackage[normalem]{ulem} 
\usepackage{upgreek} 
\usepackage{enumitem} 
\usepackage[dvipsnames]{xcolor} 
\usepackage[tbtags]{amsmath} 
\allowdisplaybreaks

\makeatletter
\def\Dated@name{}
\makeatother

\makeatletter
\def\NAT@def@citea{\def\@citea{\NAT@separator}}
\makeatother

\newcommand{\e}{\mathrm{e}}
\renewcommand{\i}{\mathrm{i}}

\begin{document}

\title{Particle acceleration in relativistic turbulence: A theoretical appraisal}

\author{Camilia Demidem} \email{demidem@apc.in2p3.fr}
\affiliation{Universit\'e  de Paris, CNRS/IN2P3, Astroparticule et Cosmologie, F-75006, Paris, France}
\affiliation{Institut d'Astrophysique de Paris, CNRS -- Sorbonne Universit\'e, \\
98 bis boulevard Arago, F-75014 Paris, France}
\affiliation{Nordita, KTH Royal Institute of Technology and Stockholm University,\\ Roslagstullsbacken 23, SE-106 91 Stockholm, Sweden\bigskip
}
\author{Martin Lemoine} 
\affiliation{Institut d'Astrophysique de Paris, CNRS -- Sorbonne Universit\'e, \\
98 bis boulevard Arago, F-75014 Paris, France}
\author{Fabien Casse}
\affiliation{Universit\'e  de Paris, CNRS/IN2P3, Astroparticule et Cosmologie, F-75006, Paris, France}

\date{Received 27 September 2019; accepted 25 May 2020; published 1 July 2020}

\begin{abstract}  
\smallskip

We discuss the physics of stochastic particle acceleration in relativistic magnetohydrodynamic (MHD) turbulence, combining numerical simulations of test-particle acceleration in synthetic wave turbulence spectra with detailed analytical estimates. In particular, we study particle acceleration in wavelike isotropic fast mode turbulence, in Alfv\'en and slow Goldreich-Sridhar type wave turbulence (properly accounting for 
anisotropy effects), including resonance broadening due to wave decay and pitch-angle randomization. At high particle rigidities, the contributions of those three modes to acceleration are comparable to within an order of magnitude, as a combination of several effects (partial disappearance of transit-time damping for fast modes, increased scattering rate for Alfv\'en and slow modes due to resonance broadening). Additionally, we provide analytical arguments regarding acceleration beyond the regime of MHD wave turbulence, addressing the issue of nonresonant acceleration in a turbulence comprised of structures rather than waves, as well as the issue of acceleration in small-scale parallel electric fields. Finally, we compare our results to the existing literature and provide ready-to-use formulas for applications to high-energy astrophysical phenomenology.\bigskip

\noindent DOI: \href{https://doi.org/10.1103/PhysRevD.102.023003}{10.1103/PhysRevD.102.023003}
\bigskip

\end{abstract}

\maketitle

\section{Introduction}
\label{sec:1}

The physical mechanisms that govern energy dissipation and conversion in astrophysical sources and the acceleration of particles, from subrelativistic momenta to the most extreme observed cosmic-ray energies, are long-standing problems in space plasma physics and high-energy astrophysics. Magnetized turbulence plays a central role in this field, whether directly or indirectly. At the least, it provides the essential scattering agent at the core of diffusive shock acceleration~\cite{87Blandford,01Malkov}, and reconnection itself is intimately linked to turbulence, either because turbulence generates reconnection, e.g., Ref.~\cite{2012SSRv..173..557L,*2017ApJ...838...91K}, or the converse~\cite{2006Natur.443..553D,*12Hoshino,*2016PhPl...23l0704D,*2019arXiv190108308G}. Electromagnetic turbulence indeed provides an efficient source of particle acceleration, following Fermi's original idea \cite{49Fermi} that a particle interacting with randomly moving magnetic mirrors of typical velocity dispersion $\beta_\text{m} c$  gains energy in a stochastic manner at a rate $\propto {\beta_\text{m}}^2c^2$. Such stochastic acceleration has been invoked to explain high-energy emission in a variety of astrophysical environments, from solar system plasmas to the remote high-energy Universe, e.g., impulsive solar flares \cite{96Larosa,*2004ApJ...610..550P,*2004MNRAS.354..870S,*2012ApJ...754..103B}, the Galactic center \cite{04Liu,*2006ApJ...647.1099L}, accretion disks \cite{98Quataert,*2018JPlPh..84c9010C,*2019MNRAS.485..163K}, pulsar wind nebulae \cite{2016JPlPh..82d6301L,*2019ApJ...872...10X}, galaxy clusters \cite{2007MNRAS.378..245B,*2008ApJ...682..175P,*2011MNRAS.410..127B,*2016MNRAS.458.2584B,*17Eckert}, active galactic nuclei \cite{18Asano}, gamma-ray bursts \cite{1996ApJ...461L..37B,*2016PhRvD..94b3005A,*2017ApJ...846L..28X}, etc.

Astrophysical collisionless turbulence is commonly described as an energy cascade spanning orders of magnitude in length scales, from a large stirring scale down to the small wavelengths of dissipative physics, with, in general, most of the fluctuation power in velocity and electromagnetic fields being carried by the larger scales. The usual order-of-magnitude timescale for particle acceleration reads $t_\text{acc}\sim t_\text{scatt}/\beta_\text{m}^2$, where $t_\text{scatt}$ represents the scattering timescale, i.e., the time needed for the particle to start diffusing in the turbulence. Hence, for the purpose of accelerating particles to high energies in cosmic plasmas, which is the main theme of this paper, one is interested in particles interacting with large-scale modes of a fast-moving turbulence spectrum: large scale, because the scattering timescale is a growing function of energy, and fast moving because of the scaling of the acceleration timescale. Here, we will thus be interested in the physics of particle interactions with relativistic ideal magnetohydrodynamic (MHD) turbulence, the ideal Ohm's law being applicable on the scales of interest. 

How particles gain energy in a turbulent setting can be addressed using a variety of theoretical or numerical tools. Quasilinear theory (QLT), e.g., Ref.~\cite{66Kennel,*67Hall,*68Lerche,*1981A&A....97..259A,*92Jaekel,*98Schlick}, provides an analytical estimate of the various diffusion tensor components to first order in the spectrum of electromagnetic fluctuations, which are commonly described as a sum of linear eigenmodes of the plasma. In the ideal MHD approximation, those are the incompressible Alfv\'en modes (thereafter indexed with $\text{A}$) as well as the fast and slow magnetosonic modes (indexed by $\text{F}$ and $\text{S}$), while entropy modes, pure density perturbations advected with the medium, do not play any role. Recent nonlinear extensions allow to include part of the perturbative expansion in turbulent fluctuations by introducing corrections to the particle trajectory, e.g., Ref.~\cite{1973Ap&SS..25..471V,*2003ApJ...590L..53M,*2008ApJ...673..942Y,*2009ASSL..362.....S} and references therein. An important outcome of quasilinear  calculations is the existence of resonant particle-wave interactions which provide possibly fast scattering rates, hence short acceleration timescales. However, whether collisionless turbulence can be realistically described as a sum of waves is a long-standing debate; see, e.g., Ref.~\cite{2014arXiv1404.2913H} for a recent appraisal. Furthermore, the intrinsic anisotropic nature of modern MHD turbulence theories~\cite{95GS,*1997ApJ...485..680G,*2000ApJ...539..273C,*01Maron,*03Cho,*2006PhRvL..96k5002B,*2011PhRvL.106g5001B} prohibits  particle-wave resonances~\cite{2000PhRvL..85.4656C,*2002PhRvL..89B1102Y,*2004A&A...420..799S}. Hence, nonresonant phenomena, corresponding to the interaction of particles with nontrivial velocity structures, are likely to play a role, and our investigation will confirm this point of view.

On the numerical front, the physics of transport and acceleration can be probed i) by following test particles in a synthetic turbulence generated from a sum of plane waves, e.g., Refs.~\cite{99Michalek,*2013ApJ...777..128L,*2019ApJ...873...13R,09Sullivan,14Fatuzzo}; ii) by following test particles in full three-dimensional (3D) MHD simulations \cite{2004ApJ...617..667D,*2006ApJ...638..811C,*2009ApJ...707..404L,*2011ApJ...728...60B,*2013ApJ...779..140X,*2014ApJ...783..143D,*2016A&A...588A..73C,14Lynn}; or iii), more recently, from 3D particle-in-cell (PIC) simulations \cite{2015PhRvL.114q5002W,17Zhdankin,*2018MNRAS.474.2514Z,*2018ApJ...867L..18Z,*18Zhdankin,18Comisso,19Wong,2019arXiv190901420C}. While the latter method offers a fully kinetic picture of the collisionless turbulence, from plasma length scales upward, thus allowing, in particular, a self-consistent treatment of the early injection and acceleration stages, MHD simulations provide a useful representation of the largest length scales with a potentially large dynamic range. Although the first method i) remains subject to the criticism of describing the turbulence as a superposition of linear waves, it allows to relate in a direct way the theoretical predictions for acceleration to the assumptions of the model and to probe effects beyond quasilinear theory. It therefore represents an interesting tool to interpret the results of more evolved simulations, which remain expensive if there are to cover a reasonable dynamic range, and are, besides, possibly sensitive to some degree, to how the turbulence is initialized. 

Hence, the above analytical and numerical methods nicely complement each other and we adopt this stance in the present paper. More specifically, we combine theoretical arguments borrowed from quasilinear and extended quasilinear theories as applied to modern turbulence theories to derive predictions for acceleration in relativistic MHD wave turbulence, which we compare to test-particle simulations. With respect to previous work on this topic, our work improves on Ref.~\cite{09Sullivan}, which considered an isotropic bath of Alfv\'en waves and neglected resonance broadening effects; it improves on Ref.~\cite{19Teraki}, which considered isotropic turbulence of pure fast, slow and Alfv\'en waves, by taking into account the effect of anisotropy and mode decay in the turbulence spectrum; it also improves on Ref.~\cite{14Fatuzzo} by paying attention to the alignment of small-scale modes to the local large-scale magnetic field, which has dramatic consequences on the acceleration physics, as we show; we also improve on the previous semianalytical studies of Refs.\cite{98Schlick,2000PhRvL..85.4656C} by considering the relativistic regime of wave phase velocities. Finally, we also provide original theoretical estimates for turbulent acceleration in the absence of resonant wave-particle interactions.

Our paper is laid out as follows. Section~\ref{sec:F} addresses the case of a pure fast mode turbulence, Sec.~\ref{sec:S} that of a pure slow mode turbulence, and Sec.~\ref{sec:A} that of pure Alfv\'en turbulence. For each, we compare analytical predictions to numerical simulations. In Sec.~\ref{sec:nr}, we discuss some aspects of stochastic acceleration outside the realm of ideal MHD wave turbulence, i.e., we provide estimates for nonresonant acceleration and characterize the influence of small-scale violations of Ohm's law. Our numerical and analytical results are brought together in Sec.~\ref{sec:disc} and compared to recent {\it ab initio} simulations of turbulent acceleration. Finally, we provide a summary in Sec.~\ref{sec:summ}. Unless otherwise noted, we assume ultrarelativistic particles and use units such that $c\,=\,1$.

\section{Fast mode turbulence}\label{sec:F}
Simulations of subrelativistic compressible MHD turbulence seem to indicate that Alfv\'en and slow magnetosonic modes are the main contributors to the kinetic and/or magnetic energy spectra, with little contribution from fast magnetosonic modes \cite{03Cho,10Kowal,17Andre}, although recent work~\cite{2019arXiv190701853M} indicates that this very result depends on how turbulence is driven at the outer scale. In any case, this does not exclude that fast modes can influence particle acceleration, depending on their scattering efficiency. Besides, simulations by \cite{16Takamoto} suggest that in the relativistic regime, fast modes are strongly coupled to Alfv\'enic modes and maintain a substantial fraction of energy, which furthermore tends to increase with magnetization.   

Given the variety of numerical setups used to simulate MHD turbulence, the size of the  parameter space to be probed (e.g., the $\beta$-parameter of the plasma, the magnetization, the strength of the turbulence, etc.) and the large uncertainty on the scaling laws obtained from simulations or observations, it is still not clear which of the available phenomenology, e.g., Refs.~\cite{41K,63I,65K,95GS,06Boldyrev}, appears best suited to describe the properties of the cascade for these different modes. Here, we rely on the general result that emerges from theoretical considerations and simulations in the subrelativistic limit, which defines the cascade of fast modes as isotropic, with a power-spectrum index around $1.5$--$1.7$, and an anisotropic cascade of Alfv\'en modes and (passive) slow modes, {\it \`a la} Goldreich-Sridhar.

\subsection{Theoretical predictions}
For fast modes, we introduce the power spectrum
\begin{equation}
\mathcal S_k^\text{F} \,=\, \frac{\eta}{1-\eta}B_0^2\,\frac{\vert 1-q_\text{F}\vert}{8\uppi k_\text{min}^{3}}\,\left(\frac{k}{k_\text{min}}\right)^{-q_\text{F}-2},
\label{eq:SkFMS}
\end{equation}
properly normalized over the range of wave vectors $k\,\in\,\left[k_\text{min},\,k_\text{max}\right]$ with $k_\text{max}\,\gg\,k_\text{min}$, to the amplitude of the magnetic perturbations contributed by fast modes, $2\int\mathrm{ d}^3k\,\mathcal S_k^\text{F}\,=\,\eta_\text{F}\,B_0^2/(1-\eta)\,=\,\eta\,B_0^2/(1-\eta)$, where the last equality holds for pure fast mode turbulence and the factor $2$ accounts for the summation over positive and negative frequencies. Here, $\eta=\langle \updelta B^2\rangle/\left(B_0^2 + \langle\updelta B^2\rangle\right)$ characterizes the relative magnitude of the turbulent magnetic energy density.

The polarization of fast modes in special relativistic MHD is given by, e.g., Ref.~\cite{*2016ApJ...827...44L,*2018MNRAS.475.2713D},
\begin{align}
\begin{split}
\omega_\text{F} &=\pm\frac{k}{\sqrt{2}} \Biggl\{ {\beta_{\text{F}\perp}}^2 +{\beta_\text{A}}^2 {\beta_\text{s}}^2{\mu_k}^2   \\
    &\quad+ \left[ \left( {\beta_{\text{F}\perp}}^2 +{\beta_\text{A}}^2 {\beta_\text{s}}^2{\mu_k}^2 \right)^2 -4 {\beta_\text{A}}^2 {\beta_\text{s}}^2{\mu_k}^2   \right]^{1/2}    \biggr \}^{1/2},
\end{split}\label{eq:o_F}\\
\updelta \boldsymbol{B_k}^\text{F} &=\updelta b_k\,\left(-\frac{k_\parallel}{k_\perp}\frac{\boldsymbol{k_\perp}}{k} +\frac{k_\perp}{k_\parallel}\frac{\boldsymbol{k_\parallel}}{k}\right)\!, \label{eq:dB_F}\\
\updelta\boldsymbol{u_k}^\text{F} &=\frac{\omega_\text{F}}{k}\,\frac{\updelta b_k}{B_0}\,\left( \frac{\boldsymbol{k}_\perp}{k_\perp} + \frac{{\beta_\text{s}}^2 k_\perp }{{\omega_\text{F}}^2-{\beta_\text{s}}^2{k_\parallel}^2} \boldsymbol{k}_\parallel\right)\!, \label{eq:du_F}\\
\updelta \boldsymbol{E_k}^\text{F} &=\frac{\omega_\text{F}}{k}\,\updelta b_k\,\frac{\boldsymbol{B_0}\times\boldsymbol{k}_\perp}{B_0 k_\perp}, \label{eq:dE_F}
\end{align}
where $\beta_\text{A}$, $\beta_\text{s}$ and $\beta_{\text{F}\perp}=\left({\beta_A}^2+{\beta_s}^2-{\beta_A}^2 {\beta_s}^2\right)^{1/2}$ denote the Alfv\'en speed, the sound speed and the fast speed, respectively. The wave vector $\boldsymbol{k}$, with modulus $k$, is decomposed over a parallel and a perpendicular component with respect to $\boldsymbol{B}_0$: $\boldsymbol{k}_\parallel=k_\parallel\boldsymbol{B}_0/B_0$, $\boldsymbol{k}_\perp=\boldsymbol{k}-\boldsymbol{k}_\parallel$, while $\mu_k\,=\,k_\parallel/k$. The (complex) perturbation amplitude $\updelta b_k$ is related to the power spectrum through $\mathcal S_k^\text{F}\,=\,\langle\left\vert\updelta b_k\right\vert^2\rangle$.

In our analytical predictions, we neglect the dependence on $\mu_k$ of the phase speed, $\beta_\text{F}$; in particular for $\beta_\text{A}\beta_\text{s}\ll \beta_{\text{F}\perp}$, we use the simplified dispersion relation $\omega_\text{F}\,\simeq\,\pm\beta_{\text{F}\perp} k$, which provides a good approximation over the range of $\mu_k\,\in\,[-1,1]$, while for $\beta_\text{A}\approx \beta_\text{s}$, the value of $\beta_\text{F}$ is set to the average over $\mu_k$ of the phase speed. These analytical predictions are obtained through quasilinear theory, which evaluates the second-order moments
$\left\langle\Delta\mu^2\right\rangle$ and $\left\langle \Delta p^2\right\rangle$ by direct integration over the electromagnetic fields that the particle experiences over a trajectory which is described by the zeroth-order gyration around the background magnetic field; see Ref.~\cite{92Jaekel} for details. Everywhere in this paper, $\mu$ represents the pitch-angle cosine of the particle with respect to the background magnetic field, and $\boldsymbol{p}$ represents its momentum; hence $\mu\,=\,p_\parallel/p$.

A detailed application of quasilinear theory then provides the following form for the diffusion coefficients:
\begin{align}
\begin{split}
\tilde{D}_{\mu\mu}^\text{F}&=\frac{\Omega^2 \uppi^2(1-\mu^2)}{2B_0^2}\sum_{\pm}\sum_{n=-\infty}^{+\infty}\int_{k_\text{min}}^{k_\text{max}} \mathrm{d}k\\ 
&\,\quad \times \int_{-1}^{1} \mathrm{d}\mu_k\,\mathcal R_{\boldsymbol{k}}\left[J_{n+1}(z_\perp)-J_{n-1}(z_\perp)\right]^2\\
&\,\quad\times\left[(k\mu_k - \mu\omega_\text{F})^2+\gamma_\mathrm{ d}^2\mu^2\omega_\text{F}^2\right]\,\mathcal S_k^\text{F},
\end{split}
\label{eq:QLFMSDmumu}\\
\begin{split}
\tilde{D}_{pp}^\text{F}&=\frac{\Omega^2\uppi^2p^2(1-\mu^2)}{2B_0^2}\sum_{\pm}\sum_{n=-\infty}^{+\infty}\int_{k_\text{min}}^{k_\text{max}} \mathrm{d}k \,\\ 
&\,\quad\times \int_{-1}^{1} \mathrm{d}\mu_k\,\mathcal R_{\boldsymbol{k}}\left[J_{n+1}(z_\perp)-J_{n-1}(z_\perp)\right]^2\\
&\,\quad\times\omega_\text{F}^2(1+\gamma_\mathrm{ d}^2)\,\mathcal S_k^\text{F}, 
\end{split}
\label{eq:QLFMSDpp}
\end{align}
with the following notations: the sum over $\pm$ sums over positive and negative real frequencies $\omega_\text{F}$, while that over $n$ runs over the harmonics of the gyrofrequency; $\gamma_\text{d}>0$ accounts for the possible finite lifetime of modes, i.e., the full mode pulsation $\omega\,=\,\omega_\text{F}\,-\,i\gamma_\text{d}\vert\omega_\text{F}\vert$; $\Omega\,=\,c/r_\text{g}\,=\,e B_0/p$ and  $z_\perp\,\equiv\,k_\perp\Omega^{-1}\sqrt{1-\mu^2}$. 

Finally, Eqs.~(\ref{eq:QLFMSDmumu}) and (\ref{eq:QLFMSDpp}) can be averaged over $\mu$ to obtain the quasilinear predictions for the diffusion coefficients for an isotropic population of particles, $D_{\mu\mu}$ and $D_{pp}$, linked by the relation
\begin{equation}
    D_{pp}\sim p^2 \beta_\text{F}^2\,D_{\mu\mu}.
\end{equation}
Below, we provide the theoretical scalings for $D_{\mu\mu}$; those for $D_{pp}$ can be directly obtained through the above relation.

The resonance function $\mathcal R_{\boldsymbol{k}}$ characterizes the interaction between the particle and a given mode. Various forms for this function in standard and extended quasilinear theories, which account for resonance broadening through wave decay and partial randomization of the particle pitch-angle cosine, are introduced and discussed in Appendix~\ref{sec:Appbrd}. By wave decay, it is meant that modes are assigned a finite lifetime, which effectively implies a temporal correlation of finite extent. Here, we do not include this effect for fast modes, but we will comment on its possible influence. By contrast, the partial randomization of the pitch-angle cosine of the particle in a turbulent bath is guaranteed, because the direction of the total magnetic field does not coincide exactly with that of $B_0$; hence $\mu$, which is defined relative to the latter, is effectively a random quantity (see Appendix~\ref{sec:Appbrd} for details).

Neglecting this effect for the time being, and considering furthermore $\gamma_\text{d}\rightarrow0$, we have $\mathcal R_{\boldsymbol{k}}=\delta\!\left(k\mu_k\mu - \omega_\text{F} + n\Omega\right)$\!,
with $n\in \mathbb{Z}$,  bringing out the infinite discrete set of resonances of standard quasilinear theory. The $n=0$ resonance describes the Landau, also called transit-time damping (TTD) resonance, which results from the interaction of the particle with the compressive modes moving along the background magnetic field~\cite{1976JGR....81.4633F,1981A&A....97..259A,98Schlick}. The $n\neq0$ resonances are described as Landau-synchrotron, or gyroresonances, between the particle motion around the background field and the motion of the mode along the field. 

For the above Dirac resonance function, we derive the following quasilinear scalings in the limit $r_\text{g}k_\text{min} \ll1$ ($r_\text{g}=c/\Omega$ the particle gyroradius),
\begin{align}
    D_{\mu\mu}^\text{F-TTD} &\sim (1-\beta_\text{F})^{\alpha}\,\frac{\eta}{1-\eta}\,\left(r_\text{g}k_\text{min}\right)^{q_\text{F}-2}\,k_\text{min},\label{eq:Dmu-F-TTD}\\
    D_{\mu\mu}^\text{F-Gyr} &\sim0.1 \frac{\eta}{1-\eta}\,\left(r_\text{g}k_\text{min}\right)^{q_\text{F}-2}\,k_\text{min} \label{eq:Dmu-F-Gyr},
\end{align}
for the quasilinear TTD and gyroresonant contributions of the pitch-angle diffusion coefficient, where $\alpha\approx 4$ is a ($\beta_\text{A}$-dependent) effective exponent. This scaling reveals that the transit-time damping contribution is strongly suppressed in the relativistic limit $\beta_\text{F}\rightarrow1$. This is easily understood from the resonance condition,  $k\mu_k\mu\, \pm\, \beta_\text{F}k\,=\,0$, which requires $\vert\mu\vert\geq\beta_\text{F}$, hence a superluminal parallel phase velocity when $\beta_\text{F}\simeq1$, thus effectively shutting down the resonance. 

Accounting for resonance broadening restores part of the transit-time damping contribution in the relativistic limit. To see this, consider the resonance broadening that results from pitch-angle randomization, with $\langle\mu\rangle\sim\langle\Delta\mu^2\rangle^{1/2}\sim\eta^{1/4}$, as discussed in Appendix~\ref{sec:Appbrd}, for strongly magnetized particles, meaning $r_\text{g}k_\text{min} \ll1$. Then $\mathcal R_{\boldsymbol{k}} \propto\exp\!\left[-\!\left(\mu_k\eta^{1/4} - \beta_\text{F}\right)^2/2\eta^{1/2}\right]$. At large values of $\beta_\text{F}$, more specifically $\beta_\text{F}> \eta^{1/4}$ where the term in the exponential cannot vanish, the resonance function can be approximated by its value at $\mu_k=1$ and a width $\beta_\text{F}/\eta^{1/4}$ in $\mu_k$. One then obtains an estimate for  $D_{\mu\mu}^\text{F-TTD}$ that is comparable to, albeit sligthly smaller than, $D_{\mu\mu}^\text{F-Gyr}$ at $\beta_\text{F}\sim1$. 

For $\beta_\text{F}<\eta^{1/4}$, the resonance $k\mu_k\langle\mu\rangle - \beta_\text{F}k\,=\,0$ can be met at values $\vert\mu_k\vert<1$; hence, one recovers the standard quasilinear result, in which the transit-time damping contribution exceeds the gyroresonant ones by about $1.5$ orders of magnitude.

For particles of large rigidity, meaning $r_\text{g}k_\text{min} \gg1$, the scattering is dominated by gyroresonances because $D_{\mu\mu}^\text{F-TTD}\sim \eta\left(r_\text{g}k_\text{min}\right)^{-3}k_\text{min}$ while 
$D_{\mu\mu}^\text{F-Gyr}\sim\eta\left(r_\text{g}k_\text{min}\right)^{-2}k_\text{min}$. The latter scaling is typical of a particle of gyroradius $r_\text{g}$ interacting with a turbulence on coherence scales $k_\text{min}^{-1}\ll r_\text{g}$.

\subsection{Numerical simulations and discussion}
Appendix~\ref{sec:num} presents in detail our numerical procedure for the Monte Carlo simulations of test-particle transport and acceleration in a given synthetic turbulence, described as a sum of linear waves. It appears important to recall here that the electric field is calculated using the ideal Ohm's law, see Eq.~(\ref{eq:ndE_F}), rather than as the sum of the $\updelta\boldsymbol{E_k}^\text{F}$ modes detailed above, in order to avoid the emergence of non-MHD effects~\cite{2006ApJ...637..322A}. 

Besides those parameters defining the turbulence, $q_\text{F}$, $\beta_\text{A}$, $\beta_\text{s}$, and $\eta$, the main (dimensionless) parameter that pilots the interaction of particles with waves is $r_\text{g}k_\text{min}$, with $r_\text{g}$ larger than the smallest wavelength of the turbulence ($r_\text{g}>k_\text{max}^{-1}$), unless specified otherwise. 
We stress that these Monte Carlo simulations are, by default, carried out in the rest frame of the unperturbed background plasma, which does not generally coincide with that of the plasma including perturbations. This means, in particular, that the turbulence generally carries a nonzero bulk velocity, hence a net bulk electric field in the simulation frame. By contrast, quasilinear predictions are (implicitly) carried out in a frame in which there is no bulk electric field. In order to gauge the influence of this difference of frames, we have carried out additional simulations for which the initial data of our Monte Carlo simulations are instead specified in the local fluid rest frame, in which the motional electric field vanishes by construction (see Appendix~\ref{sec:meas}). 
We will come back to this subtlety later.

To limit the size of parameter space, we restrict ourselves to the fiducial values: $q_\text{F}=5/3$, $\beta_\text{s}\approx10^{-2}\beta_\text{A}$, $\eta=0.3$. The value of $\beta_\text{s}$ exerts no influence on our results as long as $\beta_\text{s}\ll \beta_\text{A}$, since $\beta_\text{F} \simeq \beta_\text{A}$ in this limit. The quasilinear predictions for the diffusion coefficients directly scale with $\eta/(1-\eta)$, and we have checked that this scaling holds in our numerical simulations.

\begin{figure}
\includegraphics[width=0.48\textwidth]{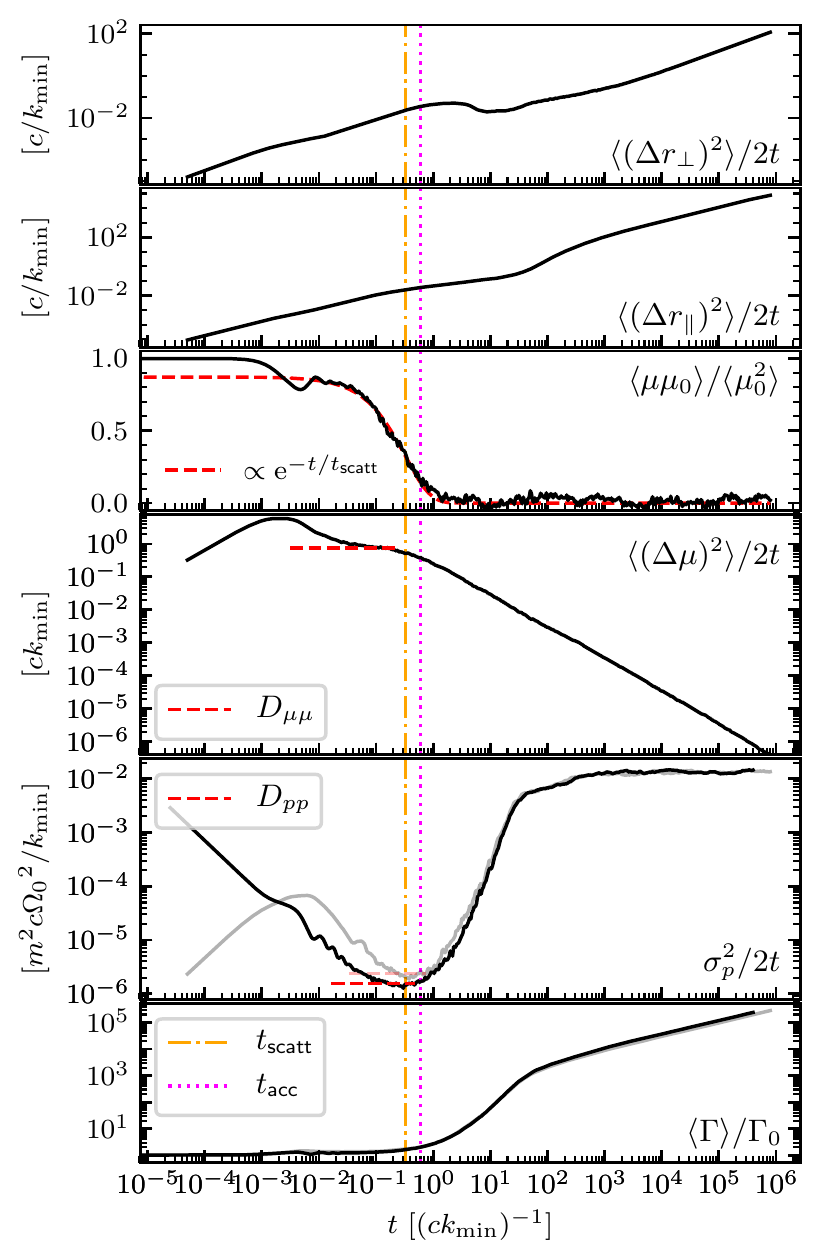}
\caption{Time evolution of various quantities for an ensemble of a thousand test particles interacting with a fast mode turbulence at $\beta_\text{A}=0.7$, $\eta=0.3$, with initially, $r_\text{g}k_\text{min}\approx 10^{-3}$. For each panel, from top to bottom: perpendicular mean square displacement; parallel mean square displacement; pitch-angle correlation function, pitch-angle mean square displacement, momentum distribution variance; finally, mean Lorentz factor normalized to its initial value. For the last two panels, the transparent (grey) and the opaque (black) curves correspond to the setups for which the rigidity is defined in the simulation frame and in the local turbulence rest frame, respectively. The red dashed lines show the fits. The orange dash-dotted line shows an estimation of the scattering timescale deduced from the fit of $\langle \mu\mu_0 \rangle$ while the magenta dotted one indicates the acceleration timescale defined as $\langle \Gamma \rangle (t_\text{acc})=2\Gamma_0$. The cyclotron frequency is written $\Omega_0=eB_0/m$.
\label{fig:rundiffF}
}
\end{figure}
We first show in Fig.~\ref{fig:rundiffF} the time evolution of various quantities for  $\beta_\text{A} = 0.7$ and $r_\text{g}k_\text{min}\approx 10^{-3}$.  This figure illustrates several noteworthy points about the acceleration of particles in a relativistic turbulence. The variance of the momentum distribution function $\sigma_p^2$ can be described, at early times, by a coherent oscillating pattern of energy gains and losses as particles gyrate around the background magnetic field and collect the influence of the perturbations. This modulation remains coherent over the coherence time of the random force that the particles suffer. Eventually, the stochastic buildup of the perturbations leads to the decorrelation of the pitch angle, after a time of the order of $t_\text{scatt}$, or around $0.3\,k_\text{min}^{-1}$ in Fig.~\ref{fig:rundiffF}. Particle acceleration occurs soon after, since the expected acceleration timescale $t_\text{acc}\sim t_\text{scatt}/\beta_\text{A}^2$. Once acceleration takes place, the behavior of $\sigma_p^2/2t$ becomes strongly superdiffusive, because the diffusion coefficient $D_{pp}$ is a growing function of energy. In detail,  quasilinear theory predicts $D_{pp}^\text{F}\propto p^{q_\text{F}}$, and the solution of the corresponding Fokker-Planck equation can be shown to exhibit in this case a momentum evolutionary law $\left\langle p^2\right\rangle\!^{1/2}\sim\left\langle p\right\rangle \propto {\Delta t}^{1/(2-q_\text{F})}\propto {\Delta t}^3$. 

Hence, once acceleration starts to operate, the particle momentum increases at a fast rate until it reaches the point where $r_\text{g}k_\text{min}\simeq1$, at $t\sim 100\,k_\text{min}^{-1}$ in Fig.~\ref{fig:rundiffF}. At larger values of the rigidity, quasilinear theory now predicts $D_{pp}^\text{F}\propto p^0$, which directly stems from the lack of resonances for  particles with gyroradius outside the inertial range of turbulence. Accordingly, at $r_\text{g}k_\text{min}\,\gg\,1$, the particle experiences a turbulence that it sees as small scale on its gyroradius scale, hence $t_\text{scatt} \propto p^2$, guaranteeing that $t_\text{acc}\propto p^2$, in agreement with $D_{pp}^\text{F}=p^2/(2t_\text{acc})\propto p^0$.
The constancy of $D_{pp}^\text{F}$ then indicates that $\sigma_p^2/(2t)$ remains fixed at this plateau value, and one recovers normal diffusion of the momentum. 

Interestingly, as the momentum evolves fast in the acceleration process, the spatial diffusion coefficient, in particular the parallel one, also becomes strongly superdiffusive. We note here that quasilinear theory cannot describe this stage, because one of its intrinsic assumption is that particles move on unperturbed orbits at constant energy, which is clearly not the case here. This super-diffusive regime has important consequences for the maximum energy of accelerated particles, because it allows them to escape faster than on a naive diffusive timescale; this is discussed in more detail in Sec.~\ref{sec:disc}.

\begin{figure}
\center \includegraphics[width=0.48\textwidth]{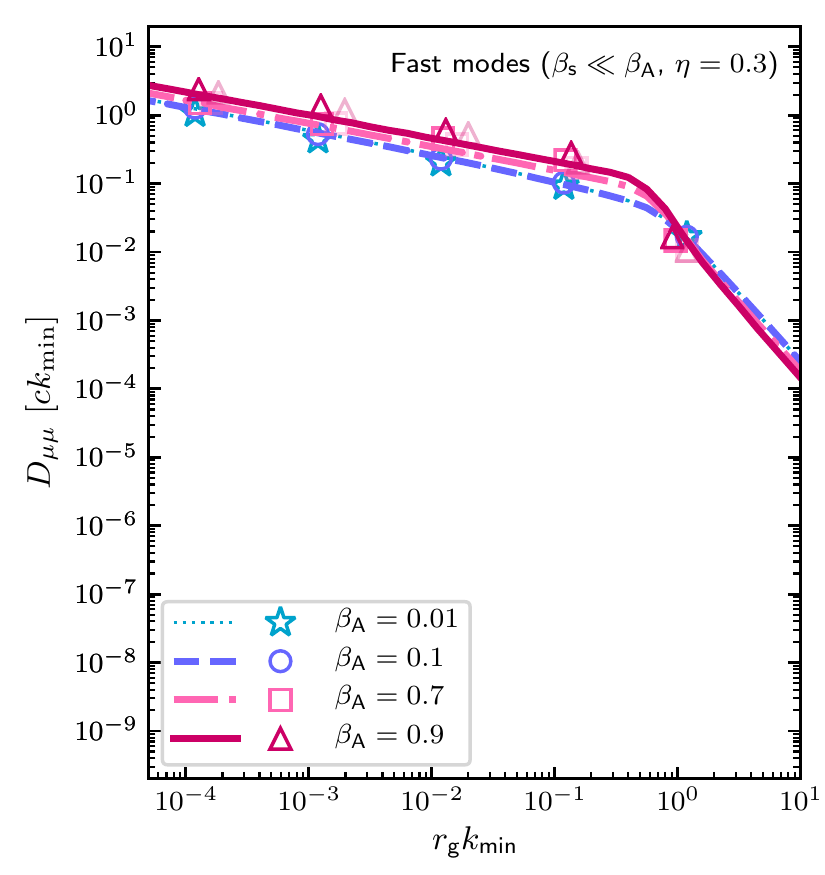}
\caption{Comparison of the theoretical QLT (with resonance broadening associated to the uncertainty in pitch angle) predictions for the $\mu$-averaged pitch-angle cosine diffusion coefficient (colored lines, for various values of $\beta_\text{A}$ as indicated) with values extracted from test-particle Monte Carlo simulations (symbols)  as a function of the (effective) initial rigidity. Following the same convention as in Fig.\ref{fig:rundiffF}, the transparent and opaque symbols refer to simulations in which the initial energy distribution of the particles is a Dirac delta function in the simulation frame and in the local turbulence rest frame, respectively. See the text for details.}
\label{plot:F_Dmu}
\end{figure}

\begin{figure}
\center \includegraphics[width=0.48\textwidth]{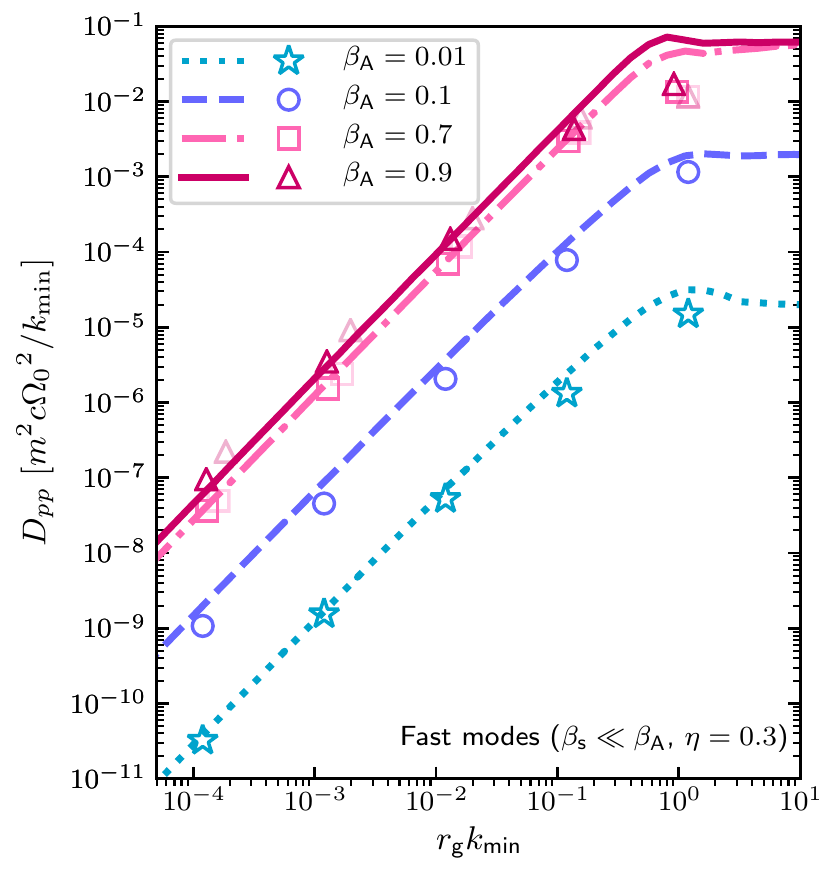}
\caption{Same as Fig.~\ref{plot:F_Dmu}, now comparing the theoretical QLT predictions for the $\mu$-averaged momentum diffusion coefficient (colored lines, for various values of $\beta_\text{A}$ as indicated) with values extracted from test-particle Monte Carlo simulations (symbols) as a function of the (effective) initial rigidity.}
\label{plot:F_Dp}
\end{figure}

We now compare the quasilinear predictions for the pitch-angle averaged diffusion coefficients $D_{\mu\mu}^\text{F}$ and $D_{pp}^\text{F}$ over a range of rigidities and mode velocities, in Fig.~\ref{plot:F_Dmu} and \ref{plot:F_Dp}. In our simulations, particles are injected with random initial directions, hence random initial pitch angles. These diffusion coefficients are thus understood as averaged over $\mu$. The details of the procedure used to derive them are given in Appendix~\ref{sec:meas}; in short, $D_{\mu\mu}$ is estimated as the slope of the linear fit of $\langle[\Delta \mu(t)]^2 \rangle/2$ while $D_{pp}$ is determined by a linear fit of $\sigma_p^2(t)/2$, as indicated by the dashed horizontal lines in the illustrative Fig.~\ref{fig:rundiffF}. Estimations of the acceleration and scattering timescales are also indicated as vertical lines, with $t_\text{acc}$ taken as the time at which the average $\left\langle p\right\rangle$ over the population of Monte Carlo particles is twice the initial value, and $t_\text{scatt}$ is defined as the decorrelation time of the pitch-angle cosine, inferred through an exponential fit of $\left\langle \mu(\Delta t)\mu(0)\right\rangle$. This procedure necessarily introduces a degree of arbitrariness, however, the scalings of the diffusion coefficients with respect to rigidity $r_\text{g}k_\text{min}$ and mode velocity $\beta_\text{A}$ (and with turbulence strength parameter $\eta$) are respected.

\begin{figure}
\center \includegraphics[width=0.48\textwidth]{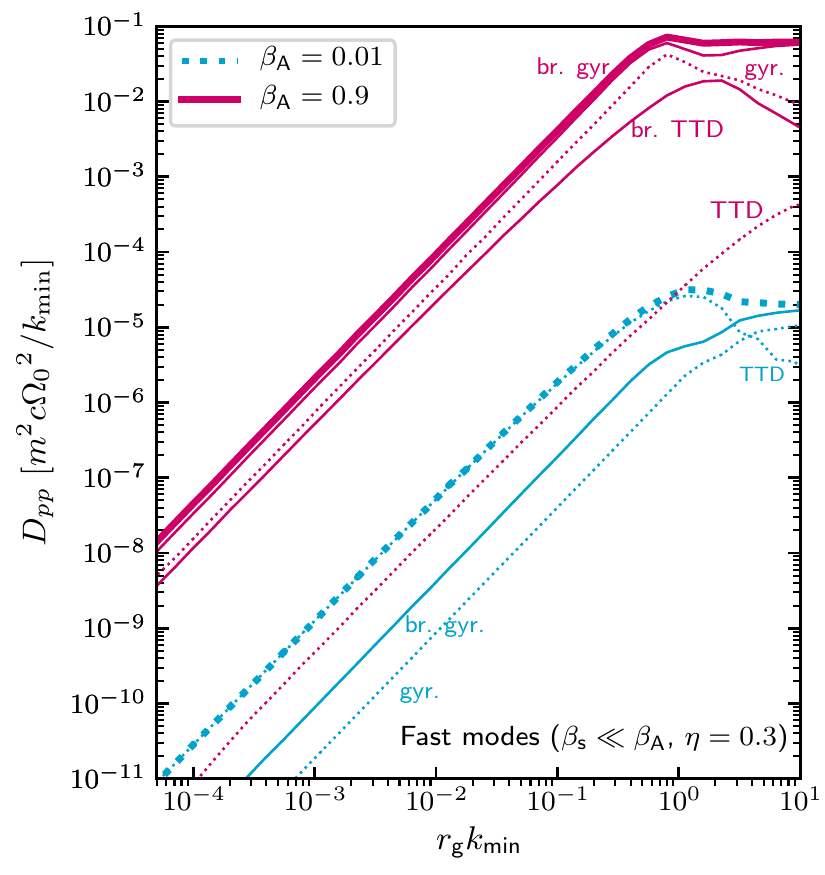}
\caption{Relative contributions to the $\mu$-averaged momentum diffusion coefficient for $\beta_\text{A}$ equal to 0.01 and 0.9. The thin dotted lines depict the linear gyroresonant and TTD contributions. The contributions accounting for resonance broadening due to the pitch-angle uncertainty are displayed in thin solid lines (br. gyr. and br. TTD). For $\beta_\text{A}=0.01$,  the curve of the broadened TTD coincides with the linear prediction and was not plotted for the sake of readability. The thick lines give the total $D_{pp}$ accounting for resonance broadening (same curves as in Fig.~\ref{plot:F_Dp}).}
\label{plot:F_Dp_contrib}
\end{figure}

As discussed above, the transit-time damping contribution largely dominates the gyroresonant contributions at low $\beta_\text{A}$, but disappears, at least partially, in the relativistic regime $\beta_\text{A}\rightarrow 1$. This feature is illustrated in Fig.~\ref{plot:F_Dp_contrib}, which shows the (theoretical) relative  contributions of TTD and gyroresonances at two extreme values of the mode velocity, $\beta_\text{A}=0.01$ and $\beta_\text{A}=0.9$. In standard quasilinear theory, the TTD contribution has almost completely disappeared at $\beta_\text{A}=0.9$. 

Here, we present the theoretical predictions for quasilinear theory including only the resonance broadening associated to the uncertainty in pitch angle because our simulations indeed consider fast modes of an infinite lifetime. Our theoretical calculations reveal that the inclusion of resonance broadening does not affect much the diffusion coefficients in fast mode turbulence, although it will play an important role in the case of slow mode anisotropic turbulence. More precisely, its effect is negligible at values $\beta_\text{A}\lesssim 0.1$, because the TTD resonance is then strongly dominant; at values $\beta_\text{A}\gtrsim 0.7$, it increases the standard QLT predictions by a factor of approximately $3$--$4$, notably due to a partial restoration of the otherwise vanishing TTD contribution (Fig.~\ref{plot:F_Dp_contrib}). Nevertheless, we wish to point out that because these predictions with resonance broadening implicitly rely on the assumption that $r_\text{g} k_\text{min}\ll 1$, they tend to overestimate the diffusion coefficients for $r_\text{g} k_\text{min}\gtrsim 1$. For these high rigidities, standard QLT is therefore better suited.

Overall, our numerical results for $D_{\mu\mu}$, and for $D_{pp}$, appear in good agreement with the theoretical expectations of extended quasilinear theory. We, however, need to come back to some corrections introduced for simulations at $\beta_\text{A}=0.7$ and even more at $\beta_\text{A}=0.9$ because of the presence of a net bulk electric field which makes the comparison with our QLT calculations less direct for the following reasons.

This electric field is of typical strength $\langle\delta E^2\rangle^{1/2}\simeq\mathcal O(\eta^{1/2}\beta_\text{A})$ and it is directly associated with a net bulk motion of the turbulence in the simulation frame, with Lorentz factor $\gamma_u \simeq \left(1+\eta u_\text{A}^2\right)^{1/2}$. When particles are initialized with a random pitch angle and a momentum $p_0$ in the simulation frame, their energy changes by $\Delta p/p_0\simeq\gamma_u^2-1$ once they interact with the turbulence, as a result of a first-order Fermi process. Consequently, their rigidity in the simulation frame is not the one that was initialized, $r_\text{g,0}k_\text{min}$, but rather $r_\text{g}k_\text{min}=[1+\Delta p/p_0] r_\text{g,0}k_\text{min}$. In the relativistic regime, this effect becomes significant for $D_{pp}$, whose scaling with $r_\text{g}$ is strong. More specifically, the values of the initial energy jump $\Delta p/p_0$ measured in the simulations are approximately $0.6$ for $\beta_\text{A}=0.9$ and $0.4$ for $\beta_\text{A}=0.7$. We used these values to report the results in Fig.~\ref{plot:F_Dp} as a function of the effective initial $r_\text{g}k_\text{min}$. These values are represented by the lighter symbols in Fig.~\ref{plot:F_Dp} and they lie above the theoretical predictions.

Given that there is no bulk electric field involved in our quasilinear computations, one can expect the local rest frame of the turbulence to be better suited to set up the initial conditions of our Monte Carlo simulations. We therefore also conducted simulations, which initialize the energy distribution as a Dirac delta function in that frame, and their results are represented by opaque symbols in Figs.~\ref{plot:F_Dmu} and \ref{plot:F_Dp}. The corresponding values of $D_{pp}$ appear to be in slightly better agreement with the quasilinear predictions, which, as mentioned earlier, also account for resonance broadening (although at high rigidities, standard QLT is better justified). \\
\indent Finally, it is noteworthy that in the relativistic limit, the $D_{pp}$ values that we measure here appear to scale as a function of the 3-velocity (rather than the 4-velocity), $D_{pp}\sim \beta_\text{F}^2$.

\section{Slow mode turbulence}\label{sec:S}

\subsection{Theoretical predictions}
\label{sec:S-th}
From a formal standpoint, the analysis for slow modes is the same as for fast modes. The expressions for the perturbation polarizations, Eqs.~(\ref{eq:dB_F})--(\ref{eq:dE_F}) remain valid, and one has simply to replace $\omega_\text{F}$ with 
\begin{equation}
\begin{split}
\omega_\text{S} &=\pm\frac{k}{\sqrt{2}} \Biggl\{ {\beta_{\text{F}\perp}}^2 +{\beta_\text{A}}^2 {\beta_\text{s}}^2{\mu_k}^2  \\
    &\quad- \left. \left[ \left( {\beta_{\text{F}\perp}}^2 +{\beta_\text{A}}^2 {\beta_\text{s}}^2{\mu_k}^2 \right)^2 -4 {\beta_\text{A}}^2 {\beta_\text{s}}^2{\mu_k}^2   \right]^{1/2}    \right \}^{1/2}\!.
    \end{split}\label{eq:o_S}
\end{equation}
Likewise, the diffusion coefficients adopt the same form as before, Eqs.~(\ref{eq:QLFMSDmumu}) and (\ref{eq:QLFMSDpp}), except that we now adopt a Goldreich-Sridhar power spectrum,

\begin{equation}
\mathcal{S}_{\boldsymbol{k}}^\text{S} =
 \frac{\eta}{1-\eta}B_0^2\, C_g \left(\frac{k_\perp}{k_\text{min}}\right)^{\!-q_\text{S}-1}g\!\left[\frac{k_\parallel}{k_\perp^{2/3}k_\text{min}^{1/3}}\right]\!,
\label{eq:GSspec}     
\end{equation}
with $g(x)$ a function peaking at $-1\lesssim x \lesssim 1$, imposing the anisotropic scaling between the parallel and perpendicular components of modes and with $C_g$, a normalization constant ensuring that $\sum_\pm \int\mathrm{ d}^3k\,\mathcal{S}_{\boldsymbol{k}}^\text{S}= \eta_\text{S} B_0^2/(1-\eta)=\eta B_0^2/(1-\eta)$. We have considered two types of anisotropy functions, $g(x)=\Theta(1-\vert x\vert)$, for which  $C_g=(q_\text{S}-1)(3q_\text{S}-5)/(16\uppi k_\text{min}^3)$, and $g(x)=\delta \! \left(\sqrt{2}x-1\right)$ for which $C_g=\sqrt{2}(3q_\text{S}-5)/(24\uppi k_\text{min}^3)$ and found that both functions lead to similar results. To simplify our analytical calculations, an approximate dispersion relation is used again, $\omega_\text{S}\approx \pm\beta_\text{S} k_\parallel$.

We now derive the predicted analytical scalings, starting with the case of standard quasilinear theory, without resonance broadening. We first recall that even though slow modes carry compressive perturbations, the TTD contribution is null in standard quasilinear theory as the linear resonance condition reads $\pm k_\parallel \beta_\text{S}-k_\parallel \mu=0$ and is virtually never met. Using Eqs.~(\ref{eq:QLFMSDmumu}) and (\ref{eq:QLFMSDpp}) and inserting the above power spectrum (with $q_\text{S}=7/3$), one derives the following scalings for gyroresonant interactions at small rigidities, i.e., $r_\text{g}k_\text{min}\ll1$,
\begin{equation}
D_{\mu\mu}^\text{S-Gyr} \sim 0.1\frac{\eta}{1-\eta} \left(r_\text{g}k_\text{min}\right)^{3/2}\! k_\text{min},
\label{eq:scalSGS}
\end{equation}
and $D_{\mu\mu}^\text{S-Gyr} \sim 0.1\eta \left(r_\text{g}k_\text{min}\right)^{-2}\,k_\text{min}$ as usual, in the high-energy limit $r_\text{g}k_\text{min}\gg1$. The strong scaling $D_{\mu\mu}^\text{S-Gyr}\propto \left(r_\text{g}k_\text{min}\right)^{3/2}$ stems from the anisotropy of the Goldreich-Sridhar spectrum~\cite{2000PhRvL..85.4656C}, since $k_\perp \gg k_\parallel$ implies that resonant particles with $r_\text{g}k_\parallel \sim 1$ will cross many uncorrelated fluctuations in the perpendicular direction. 

We now turn to the more physically motivated case of damped modes. The condition of critical balance underlying Goldreich-Sridhar model implies that the timescale of nonlinear interactions coincides with the linear propagation timescale; accordingly, we set $\gamma_\text{d}=1$ in numerical applications. In the presence of time decorrelation, the resonance function takes on a Breit-Wigner form; see Eq.~(\ref{eq:brdom}) with $\Re\omega=\omega_\text{S}$. A direct calculation of the integrals then provides the following scalings in the low-rigidity limit $r_\text{g}k_\text{min}\ll1$:
\begin{align}
D_{\mu\mu}^\text{S-TTD} &\sim0.1\frac{\eta}{1-\eta} \gamma_\text{d}\beta_\text{S}\left[1 - 3\ln \!\left(r_\text{g}k_\text{min}\right)^{}\right]\! k_\text{min},\label{eq:scalSDGS-TTD} \\ 
D_{\mu\mu}^\text{S-Gyr} &\sim0.1\frac{\eta}{1-\eta} \gamma_\text{d}\beta_\text{S} k_\text{min}.
\label{eq:scalSDGS}
\end{align}

In agreement with previous studies carried out in the nonrelativistic limit \cite{2000PhRvL..85.4656C,14Lynn,18Xu}, we find that the resonance broadening allows TTD interactions, which now dominate the transport, and provide for $q_\text{S}=7/3$, an interaction time with a weak logarithmic dependence on the rigidity\footnote{Ref.~\cite{2000PhRvL..85.4656C}, which uses an exponential decay model for the turbulence decorrelation similar to ours, obtains the same logarithmic dependence for the diffusion coefficients; our numerical values are, however, larger by about an order of magnitude.} for $r_\text{g}k_\text{min}\ll1$.  We also note that $D_{\mu\mu}^\text{S}$ scales with $\beta_\text{S}\gamma_\text{d}$ which characterizes the scaling of the decay coefficient.

In the high-rigidity limit, meaning $r_\text{g}k_\text{min}\gg1$, $D_{\mu\mu}^\text{S-TTD}  \sim\eta \gamma_\text{d}\beta_\text{S}\left(r_\text{g}k_\text{min}\right)^{-3} k_\text{min}$ while $D_{\mu\mu}^\text{S-Gyr}  \sim 0.1\,\eta \left(r_\text{g}k_\text{min}\right)^{-2} k_\text{min}$.  

Consider now the effect of the partial randomization of the pitch angle of the particle with respect to the direction of the magnetic field, as discussed in Appendix~\ref{sec:Appbrd}. To distinguish this effect from that associated to the finite lifetime of the modes, we assume here $\gamma_\text{d}=0$. Using Eq.~(\ref{eq:brdmu}),  we thus derive, for $r_\text{g}k_\text{min}\ll1$,
\begin{equation}
D_{\mu\mu}^\text{S-TTD} \sim 0.1\kappa(\eta,\beta_\text{S})\,\eta \,k_\text{min}\,\left[1 - 1.8\ln \!\left(r_\text{g}k_\text{min}\right)\right]\!,
\label{eq:scalSmuGS}
\end{equation}
with $\kappa(\eta,\beta_\text{S})\sim\exp \!\left[-(1-\beta_\text{S}/\eta^{1/4})^2\right]$ a function of $\eta$ and $\beta_\text{S}$ that stems from resonance broadening, and which is such that $\kappa\sim\mathcal O(1)$ for $\beta_\text{S}\lesssim\eta^{1/4}$. In the opposite limit, $\kappa$ drops exponentially fast toward zero as $\eta^{1/4}/\beta_\text{S}\rightarrow 0$, in agreement with the vanishing TTD contribution for slow modes in the absence of resonance broadening. 

The gyroresonant contribution is here strongly suppressed relative to its TTD counterpart, and can be neglected safely. This difference, with respect to resonance broadening by wave decay, Eqs.~(\ref{eq:scalSDGS-TTD}) and (\ref{eq:scalSDGS}), results from the shape of the resonance function, which is exponential in the present case, vs Breit-Wigner in the previous case. One can indeed show that the contribution to $D_{\mu\mu}^\text{S-Gyr}$ in Eq.~(\ref{eq:scalSDGS}) results from modes with 
$k_\perp\lesssim k_\text{min}\left(r_\text{g}k_\text{min}\right)^{-3/2}$, i.e., $k_\parallel r_\text{g}\lesssim 1$. However, imposing this constraint in the resonance function given by Eq.~(\ref{eq:brdmu}) results in an exponential suppression of the form $\exp \!\left\{-\left[n/(k_\parallel r_\text{g}\eta^{1/4})-\ldots\right]^2\right\}$, where $\ldots$ represent terms of the order of unity, and $n\geq 1$ for gyroresonant interactions.

At rigidities below the inertial range, i.e., such that $r_\text{g} k_\text{max}<1$, the $[1-1.8\ln(r_\text{g}k_\text{min})]$ prefactor is replaced by $\ln \!\left(k_\text{max}/k_\text{min}\right)$, and the $D_{\mu\mu}$ tends to a constant, as observed in Ref.~\cite{19Teraki}. 
In the high-rigidity limit, $r_\text{g}k_\text{min} \gg 1$, $D_{\mu\mu}^\text{S-TTD} \,\sim\,\kappa(\eta,\beta_\text{S}) \eta \left(r_\text{g}k_\text{min}\right)^{-3} k_\text{min}$.

\subsection{Numerical simulations and discussion}
Again, we relegate the details of the numerical setup to Appendix~\ref{sec:num}. However, we emphasize here that setting up the Goldreich-Sridhar phenomenology is not a trivial task as one must pay attention to the notion of a local mean field ~\cite{2000ApJ...539..273C}; the direction of anisotropy, which establishes the hierarchy between $k_\parallel$ and $k_\perp$, must be defined relative to the local mean field, which is the total field at the given point averaged over scales larger than the particle gyroradius. In order to account for this effect, we artificially reduce the amplitude of the turbulence down to low values, $\eta\sim0.01$, which guarantees that everywhere, the field line direction does not depart from that of $\boldsymbol{B_0}$  by an angle larger than $\updelta B/B_0\sim\eta^{1/2}$. Then, provided that the eddy anisotropy verifies $k_\parallel/k\gtrsim \updelta B/B_0$, this eddy can be considered as aligned with respect to the local mean field $\boldsymbol{B}$, even if in practice, it is aligned along $\boldsymbol{B_0}$. The interest of setting the parallel direction along $\boldsymbol{B_0}$ is, of course,  to preserve a standard Fourier decomposition throughout space for our numerical simulations. Since particle-wave resonances require $k_\parallel r_\text{g}\sim1$, with $k_\parallel/k\sim k_\parallel^{-1/2}k_\text{min}^{1/2}$ (critical balance), the above constraints now imply $r_\text{g}k_\text{min}\gtrsim \eta$.

The above setup with $\eta=0.01$ thus allows us to probe the physics of particle acceleration in a realistic anisotropic Goldreich-Sridhar-like configuration for particles with $r_\text{g}k_\text{min}\gtrsim 0.01$. Since the analytical scalings predict $D_{\mu\mu}\propto\eta/(1-\eta)$ and $D_{pp}\propto\eta/(1-\eta)$, we may, in turn, extrapolate these scalings to the regime of larger $\eta$ provided they match the numerical simulations. To ensure that $\beta_\text{S}$ remains close to unity, we consider a hot plasma with $\beta_\text{s}=1/\sqrt{3}$ and $\beta_\text{A}=0.5$. For $\beta_\text{s}\ll\beta_\text{A}$ or $\beta_\text{A}\ll\beta_\text{s}$, we have indeed $\beta_\text{S}\simeq\text{min}\left(\beta_\text{s},\,\beta_\text{A}\right)$. 

\begin{figure}
\center \includegraphics[scale=1.]{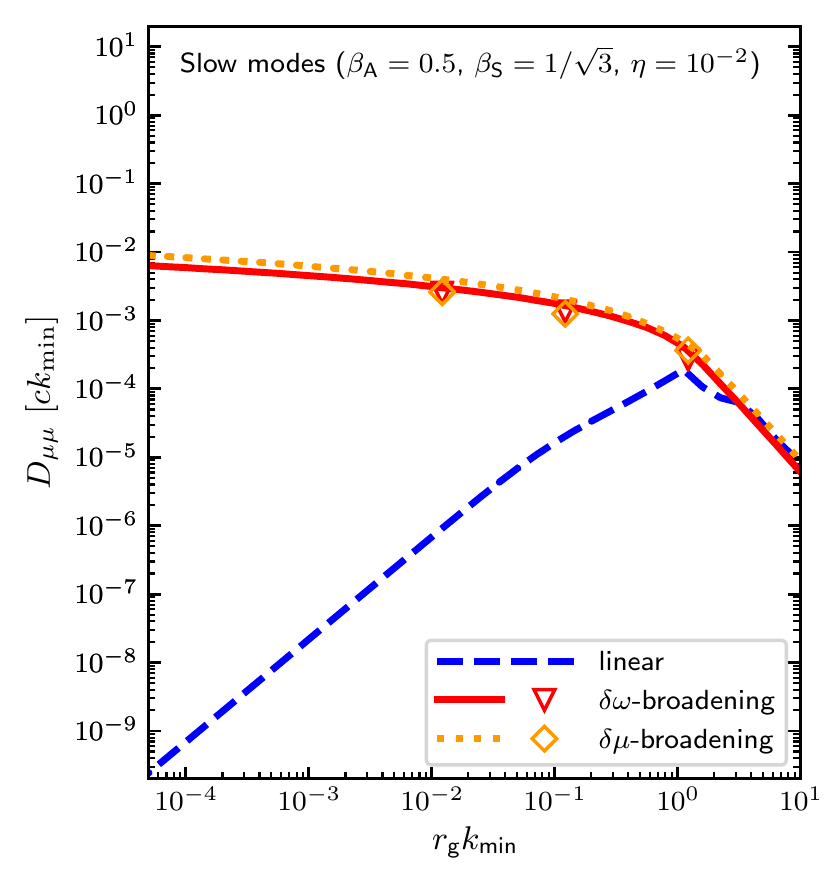}
\caption{Quasilinear predictions for the $\mu$-averaged pitch-angle cosine diffusion coefficient in a Goldreich-Sridhar-like turbulence of slow modes (lines) and corresponding values extracted from test-particle Monte Carlo simulations (symbols), plotted as a function of the initial rigidity. The legends have the following meanings: ``$\delta \omega$-broadening'' incorporates resonance broadening associated to the finite lifetime of the modes (with $\gamma_\text{d}=1$); ``$\delta \mu$-broadening'' takes into account the partial randomization of the pitch angle, while ``linear'' corresponds to the standard QLT prediction without resonance broadening. See the text for details.}
\label{plot:SGS_Dmu}
\end{figure}

\begin{figure}
\center \includegraphics[scale=1.]{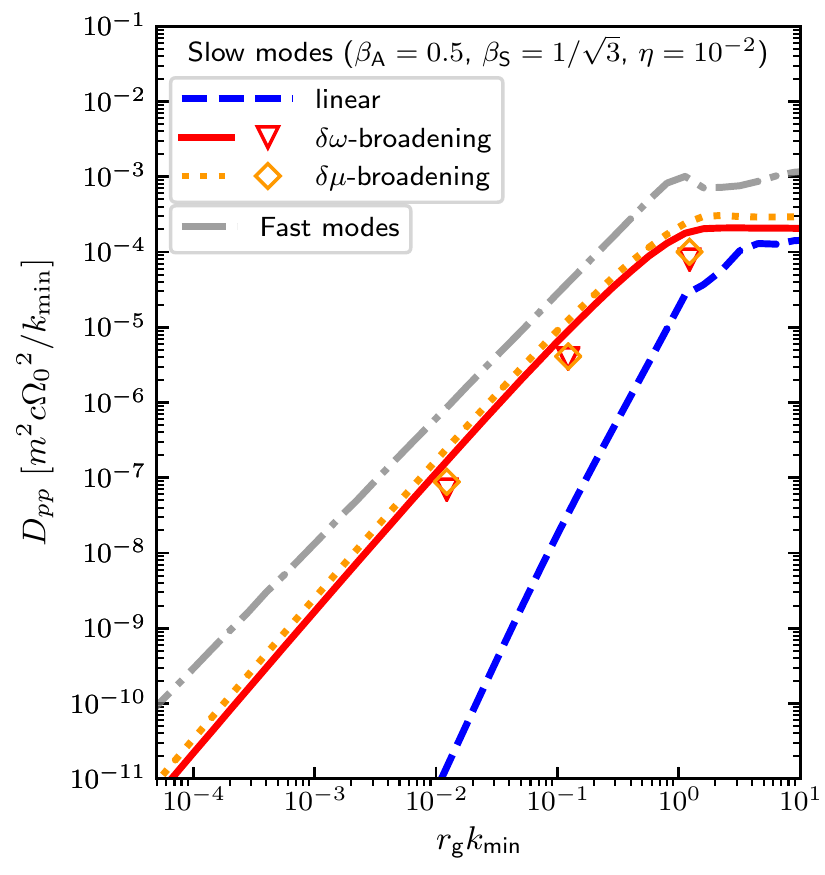}
\caption{Same as Fig.~\ref{plot:SGS_Dmu} for the momentum diffusion coefficient. For comparison, the prediction for the diffusion coefficient in an isotropic turbulence of fast modes (with resonance broadening) at same $\beta_\text{A}$, $\beta_\text{s}$ (corresponding to $\beta_\text{F}\simeq0.68$) and $\eta$ is overlaid in gray/dot-dashed.}
\label{plot:SGS_Dp}
\end{figure}

Figures~\ref{plot:SGS_Dmu} and \ref{plot:SGS_Dp} compare the generalized quasilinear predictions for the pitch-angle averaged diffusion coefficients described in Sec.~\ref{sec:S-th} to the values derived from the Monte Carlo simulations. For dynamical turbulence ($\gamma_\text{d}=1$), we find good agreement between the simulations (red symbols in the figures) and the semianalytical predictions, with $D_{\mu\mu}\sim{\rm cst}$ up to a logarithmic correction $\ln \!\left(r_\text{g}k_\text{min}\right)$.

For undamped modes ($\gamma_\text{d}=0$), however, the scattering time in Monte Carlo simulations (orange symbols) remains comparable to that for damped modes and several orders of magnitude larger than the standard quasilinear estimates (blue curves), which furthermore predict $D_{\mu\mu}\propto (r_\text{g}k_\text{min})^{3/2}$. The momentum diffusion coefficient $D_{pp}$ reveals a similar discrepancy. These results are, however, in satisfactory agreement with our theoretical predictions given in Eq.~(\ref{eq:scalSmuGS}), which include resonance broadening by pitch-angle randomization. Although this effect lies beyond quasilinear theory, which considers unperturbed orbits, such partial randomization is inherent to our numerical Monte Carlo simulations, because they follow the trajectories of the particles in the exact electromagnetic field configuration. Our test-particle simulations also indicate that the diffusion coefficient depends on the pitch-angle cosine, being more pronounced at values $\mu\simeq\pm0.5$ than at $\mu\simeq\pm1$, as expected from the broadening of the TTD resonance described above. This suggests that our theoretical scaling given in Eq.~(\ref{eq:scalSmuGS}) captures the physics that the test-particle simulations reproduce. 

Note the scale in the diffusion coefficients, in particular that of $D_{\mu\mu}$, which suggests $t_\text{scatt}\sim10^2k_\text{min}^{-1}$ at $r_\text{g}k_\text{min}\ll1$. This low value of $D_{\mu\mu}$ (equivalently, this large value of $t_\text{scatt}$) is a direct consequence of our choice $\eta=0.01$, since $D_{\mu\mu}\propto\eta/(1-\eta)$ (hence $t_\text{scatt}\propto\eta^{-1}$ for $\eta\ll 1$); see Eqs.~(\ref{eq:scalSGS}), (\ref{eq:scalSDGS-TTD}) and (\ref{eq:scalSmuGS}). Hence, extrapolating to $\eta\sim1$ predicts a scattering timescale of the order of $k_\text{min}^{-1}$. When adapting the values of $D_{pp}$, one must also keep in mind that those shown in Fig.~\ref{plot:SGS_Dp} assume $\eta=0.01$ and that $D_{pp}\propto \eta/(1-\eta)$.

\section{Alfv\'en mode turbulence}\label{sec:A}

\subsection{Theoretical predictions}
We  use the same Goldreich-Sridhar power spectrum as for slow modes, Eq.~(\ref{eq:GSspec}), and repeat the generalized quasilinear calculations of Sec.~\ref{sec:S-th} with dispersion and polarization relations appropriate for Alfv\'en modes,
\begin{align}
    \omega_\text{A} &=\pm \beta_\text{A} k_\parallel, \label{eq:o_A} \\
    \updelta \boldsymbol{B}_{\boldsymbol{k}}^\text{A} &= \updelta b_k \frac{\boldsymbol{B}_0 \times \boldsymbol{k}_\perp}{B_0 k_\perp}, \label{eq:dB_A}\\
    \updelta \boldsymbol{u}_{\boldsymbol{k}}^\text{A} &=\mp \beta_\text{A}\frac{\updelta \boldsymbol{B}_{\boldsymbol{k}}^\text{A}}{B_0},\label{eq:du_A}\\
    \updelta \boldsymbol{E}_{\boldsymbol{k}}^\text{A} &=\mp \beta_\text{A}\updelta b_k\frac{\boldsymbol{k}_\perp}{k_\perp},\label{eq:dE_A}
\end{align}
to obtain
\begin{align}
\begin{split}
       \tilde{D}_{\mu\mu}^\text{A}  &=\frac{\Omega^2\uppi^2(1-\mu^2)}{2B_0^2}\sum_{\pm}\sum_{n=-\infty}^{+\infty} \iint \mathrm{d} k_\perp \mathrm{d} k_\parallel \, k_\perp\\
        &\,\quad\times\mathcal{R}_{\boldsymbol{k}} \left[J_{n+1}(z_\perp)+J_{n-1}(z_\perp)\right]^2 k_\parallel^{-2}\\
       &\,\quad\times\left[(k_\parallel - \mu\omega_\text{A})^2+\mu^2\gamma_\text{d}^2\omega_\text{A}^2\right]\,\mathcal S_{\boldsymbol{k}}^\text{A},
       \end{split}
\label{eq:QLAGSDmumu}\\
\begin{split}
        \tilde{D}_{pp}^\text{A}&=\frac{\Omega^2\uppi^2p^2(1-\mu^2)}{2B_0^2}\sum_{\pm}\sum_{n=-\infty}^{+\infty}\iint \mathrm{d} k_\perp \mathrm{d} k_\parallel \, k_\perp  \\
        &\,\quad\times\,\mathcal{R}_{\boldsymbol{k}} \left[J_{n+1}(z_\perp)+J_{n-1}(z_\perp)\right]^2k_\parallel^{-2}\\
        &\,\quad\times\omega_\text{A}^2(1+\gamma_\text{d}^2)\,\mathcal S_{\boldsymbol{k}}^\text{A}.
        \end{split}
\label{eq:QLAGSSDpp}
\end{align}
The expressions differ from those derived for magnetosonic modes through the relative sign of the Bessel functions, which incidentally ensures that the $n=0$ term is null ($J_{-1}=-J_1$), as well as through the dispersion relation and power spectrum of the modes (which differs from that of fast modes), of course. But besides the fact that there are no TTD interactions, as a result of  the absence of compressible magnetic perturbations, the general remarks made for slow modes in Sec.~\ref{sec:S-th} remain true.

For the idealized case of undamped waves, $\gamma_\text{d}=0$, we derive the following theoretical scaling for $r_\text{g}k_\text{min}\ll1$,
\begin{equation}
    D_{\mu\mu}^\text{A-Gyr} \sim 0.1\frac{\eta}{1-\eta} \left(r_\text{g}k_\text{min}\right)^{3/2}\! k_\text{min},
    \label{eq:scalAGS}
\end{equation}
similar to that obtained for slow mode waves in the same case. At high rigidities, $r_\text{g}k_\text{min}\gg1$, $D_{\mu\mu}^\text{Gyr} \sim 0.1\eta \left(r_\text{g}k_\text{min}\right)^{-2}\! k_\text{min}$, as always.

If one now accounts for a finite lifetime of the waves, we obtain, for $r_\text{g}k_\text{min}\ll1$,
\begin{equation}
D_{\mu\mu}^\text{A-Gyr} \sim 0.1 \frac{\eta}{1-\eta} \gamma_\text{d}\beta_\text{A}\left[1 - 2.7\ln \!\left(r_\text{g}k_\text{min}\right)^{}\right]\! k_\text{min},
\label{eq:scalADGS}
\end{equation}
and that for $r_\text{g}k_\text{min}\gg1$ remains unchanged.

By contrast with slow modes, the resonance broadening that results from the partial randomization of the pitch angle does not play any significant role here, for two essential reasons. First, Alfv\'en waves do not possess a magnetic perturbation parallel to the mean magnetic field, so that the broadening is of relative order $\eta^{1/2}$, instead of $\eta^{1/4}$ for magnetosonic modes. Consequently, the resonance remains narrow if $\eta\ll1$. Second, as discussed in the case of slow modes, this source of broadening does not affect as strongly gyroresonant interactions as it does enhance TTD contributions. For this reason, we do not anticipate any effect from the partial randomization of the pitch angle, and our numerical simulations will confirm this result.

\subsection{Numerical simulations and discussion}
\begin{figure}
\center \includegraphics[scale=1.]{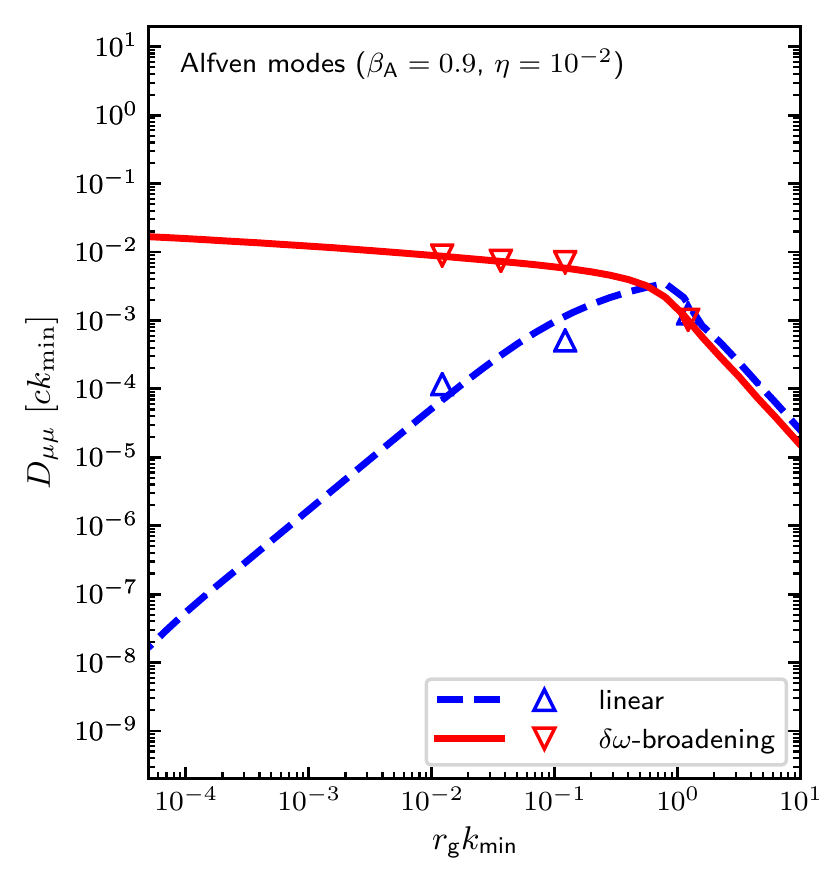}
\caption{Quasilinear predictions for the $\mu$-averaged pitch-angle cosine diffusion coefficient in a Goldreich-Sridhar-like turbulence of Alfv\'en modes, with (``$\delta \omega$-broadening'') and without (``linear'') wave decay (lines) and corresponding values extracted from test-particle Monte Carlo simulations (symbols), plotted as a function of the initial rigidity. The turbulence amplitude is here set to $\eta=0.01$. See the text for details.}
\label{plot:AGS_Dmu}
\end{figure}
\begin{figure}
\center \includegraphics[scale=1.]{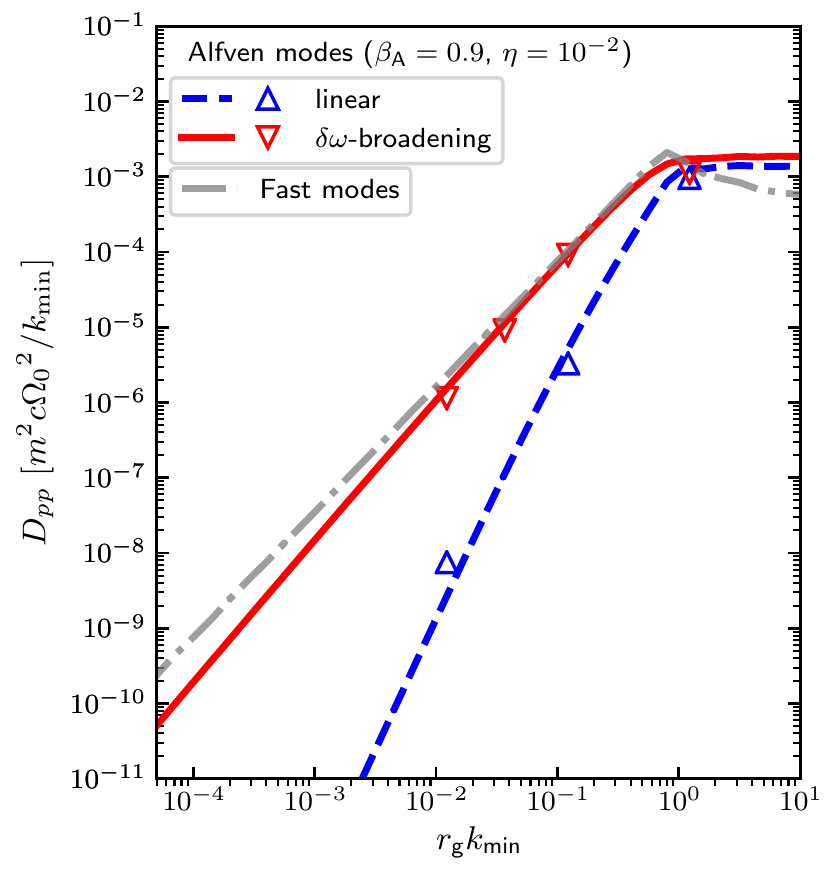}
\caption{Same as Fig.~\ref{plot:AGS_Dmu} for the momentum diffusion coefficient. For comparison, the prediction for the diffusion coefficient in an isotropic turbulence of fast modes (with resonance broadening) at same $\eta$ and $\beta_\text{A}$ and $\beta_\text{s}\ll \beta_\text{A}$ (corresponding to $\beta_\text{F}\simeq0.9$) is overlaid in gray/dot-dashed.}
\label{plot:AGS_Dp}
\end{figure}

\begin{figure}
\center \includegraphics[scale=1.]{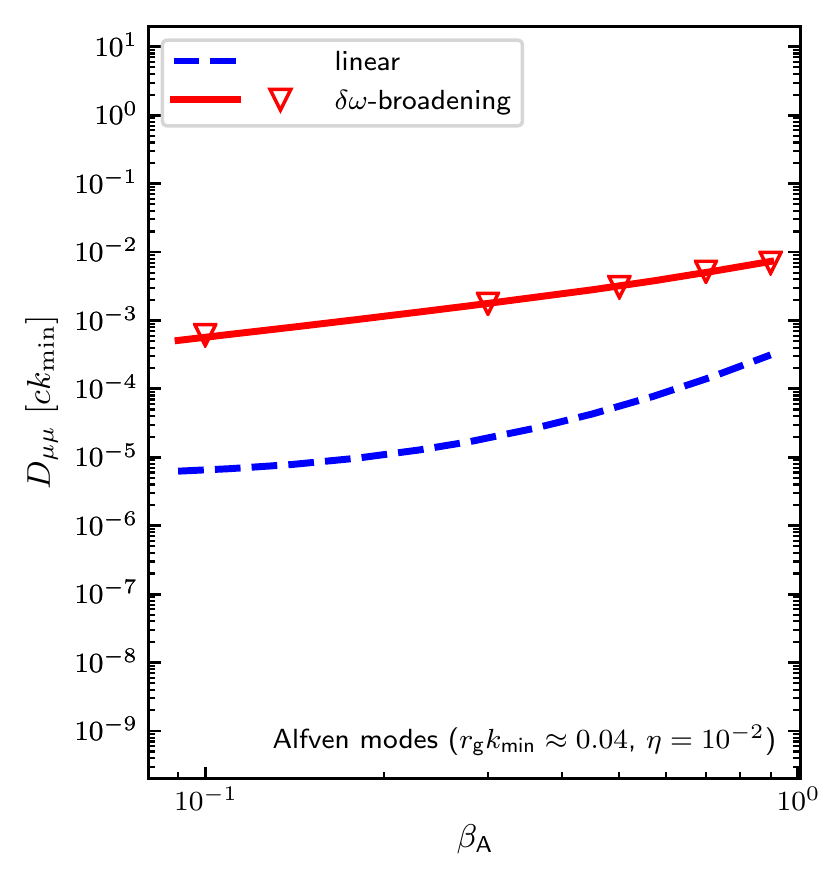}
\caption{Quasilinear predictions for the $\mu$-averaged pitch-angle cosine diffusion coefficient in a Goldreich-Sridhar-like turbulence of Alfv\'en modes with (``$\delta \omega$-broadening'') wave decay (red line) and corresponding values extracted from test-particle Monte Carlo simulations (symbols), plotted as a function of the Alfv\'en speed for a given rigidity. For reference, the prediction for the linear case is also indicated. See the text for details.}
\label{plot:AGS_vaDmu}
\end{figure}
\begin{figure}
\center \includegraphics[scale=1.]{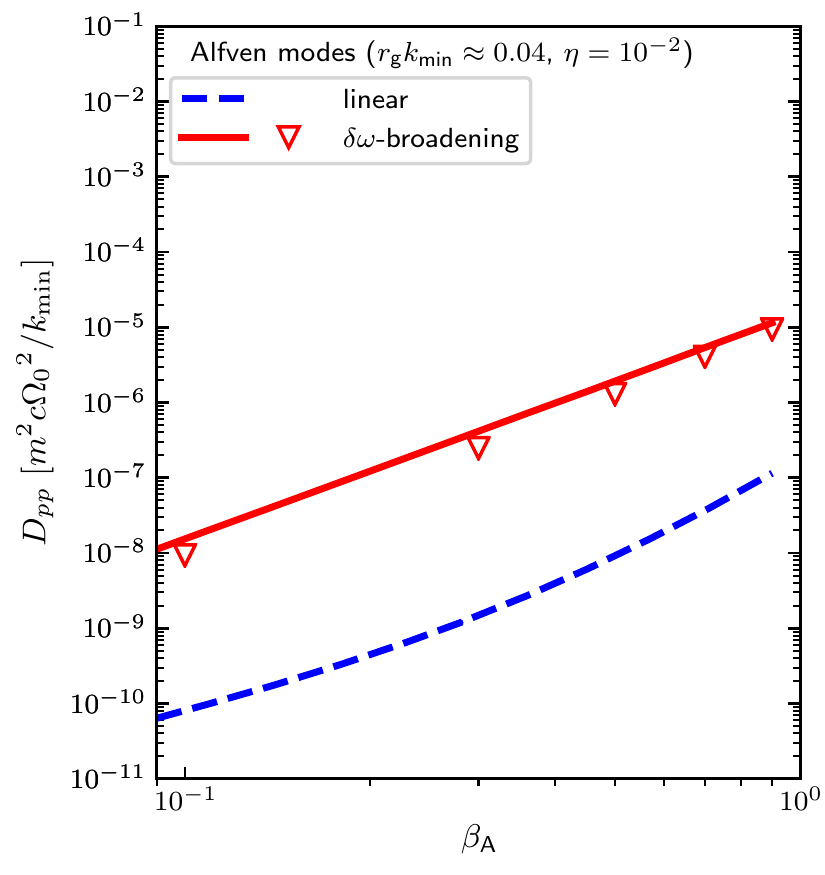}
\caption{Same as Fig.~\ref{plot:AGS_vaDmu} for the momentum diffusion coefficient.}
\label{plot:AGS_vaDp}
\end{figure}

In our Monte Carlo simulations of Alfv\'en turbulence, the Goldreich-Sridhar local anisotropy is enforced using the same technique as for slow modes: we artificially reduce the amplitude of the turbulence down to a low level, $\eta=0.01$, which allows us to probe gyroresonant interactions down to $r_\text{g}k_\text{min}\simeq0.01$.

Figures~\ref{plot:AGS_Dmu} and \ref{plot:AGS_Dp} compare the values of the diffusion coefficients obtained from numerical estimations of Eqs.~(\ref{eq:QLAGSDmumu}) and (\ref{eq:QLAGSSDpp}) to those derived from our Monte Carlo simulations for the fiducial parameters $\eta=10^{-2}$, $\beta_\text{A}=0.9$.

In these figures, the blue symbols (respectively, the blue dashed lines) correspond to the Monte Carlo simulation results (respectively, the quasilinear predictions) for undamped waves. They are found to agree fairly well one with the other, thus confirming the theoretical scalings of this (idealized) Goldreich-Sridhar phenomenology. We note, in particular, the difference with respect to slow mode waves, for which a non-negligible scattering frequency was observed in our test-particle Monte Carlo simulations for this case, because of the broadening of the TTD contribution by pitch-angle randomization.

For damped modes, we find again reasonable agreement for the $r_\text{g}k_\text{min}$-scaling, while Figs.~\ref{plot:AGS_vaDmu} and \ref{plot:AGS_vaDp} support the $\propto \beta_\text{A}$ and $\propto \beta_\text{A}^3$ scaling of $D_{\mu\mu}$ and $D_{pp}$, respectively.

Finally, we note here as well that the values of $D_{\mu\mu}$ and $D_{pp}$ take comparatively low values because of our choice $\eta=0.01$. To extrapolate these values to larger values of $\eta$, one must keep in mind that both scale linearly with $\eta/(1-\eta)$; see Eqs.~(\ref{eq:scalAGS}) and (\ref{eq:scalADGS}). In particular, for $\eta\sim1$, one expects a scattering timescale of the order of $k_\text{min}^{-1}$ at $r_\text{g}k_\text{min}\ll1$.

\section{Beyond MHD wave turbulence}\label{sec:nr}

\subsection{Nonresonant turbulent acceleration}
\label{sec:nra}
So far, we have discussed particle acceleration through resonant (or quasiresonant) interactions with MHD waves. Yet, it is not obvious that such linear eigenmodes of the plasma provide a faithful description of actual modes in strongly interacting turbulence. In this section, we address some salient features of the nonresonant acceleration of a particle in a (relativistic) turbulent bath.

On a general level, this physics can be described as the interaction of a particle with a fluctuating electric field that is directly associated through ideal Ohm's law to the fluctuating velocity field of the plasma. Particles then gain energy as they experience the compressive, shearing, accelerating and vortical motions of the medium, e.g.~\cite{1983ICRC....9..313B,*1988SvAL...14..255P,*2003ApJ...595..195W,*2010ApJ...713..475J,2013ApJ...767L..16O,2019PhRvD..99h3006L}. 

To characterize the acceleration rate, one needs to specify the statistics of the velocity field and how the particle experiences the resulting electric fields, i.e., how it is transported across the cells of coherence of the turbulence. At large rigidities, meaning $r_\text{g}k_\text{min} \gg 1$, particles cross a coherence cell in a near-ballistic manner; hence, the scattering timescale can be directly expressed as $t_\text{scatt} \simeq \eta^{-1}\,k_\text{min}^{-1}\left(r_\text{g}k_\text{min}\right)^2$. For magnetized particles, i.e., $r_\text{g}k_\text{min} \ll 1$, the problem is, however, much more complex and lies beyond the scope of the present study. Here, we rather aim to discuss how a given scattering timescale $t_\text{scatt}$ impacts the acceleration process. This scattering timescale $t_\text{scatt}$ may be smaller or larger than $k_\text{min}^{-1}$, with important consequences for the physics of acceleration.

A key point, indeed, is that the particle experiences in a different way the modes on large scales $k^{-1}$, i.e., such that $t_\text{scatt} \lesssim k^{-1}$, and the small-scale modes for which $k^{-1}\lesssim t_\text{scatt}$~\cite{2019PhRvD..99h3006L}. Specifically, for a turbulent spectrum strongly peaked on a single scale $K^{-1}$, the diffusion coefficient obeys
\begin{align}
D_{pp}& \simeq p^2\left\langle (\partial u)^2\right\rangle_K t_\text{scatt} &\left(t_\text{scatt}K\ll1\right)\,\nonumber\\
&\simeq p^2\frac{\left\langle (\partial u)^2\right\rangle_K}{K^2 t_\text{scatt}} &\,\left(t_\text{scatt}K\gg1\right)\!.
\label{eq:scalD}
\end{align}
In the above formula, the quantity $\left\langle (\partial u)^2\right\rangle_K$ represents, in a symbolic way, the contribution of the compressive, shearing, accelerating and vortical motions on scale $K^{-1}$, to the acceleration process. In detail, the decomposition of the random fluid 4-velocity in terms of acceleration (characterized by a 3-vector $\boldsymbol{a}$), compressive (characterized by a compression scalar $\theta$), shearing (3-tensor $\sigma^{ij}$), and vortical (3-tensor $\omega^{ij}$) motions leads to
\begin{equation}
    \partial_\alpha\updelta u_i = {\delta^0}_\alpha\,a_i+{\delta^j}_\alpha\left( \frac{1}{3}\,\theta\delta_{ij}+\sigma_{ji}+\omega_{ji}\right)\!,
\label{eq:helmu}
\end{equation}
with: $\theta = \boldsymbol{\nabla}\cdot\updelta\boldsymbol{u}$, $\sigma_{ij} = (\partial_i u_j + \partial_j u_i)/2-\frac{1}{3}\,\delta_{ij}\,\theta$, and $\omega_{ij} = (\partial_i u_j - \partial_j u_i)/2$, $i,j\in\left\{1,2,3\right\}$ and $\alpha\in\left\{0,1,2,3\right\}$. In the subrelativistic limit $\left\langle\updelta u^2\right\rangle \ll 1$ ($\updelta u$ is here understood as a 4-velocity), only the compressive and shearing terms contribute to leading order. In the highly relativistic limit, however, all terms contribute roughly equally; see Ref.~\cite{2019PhRvD..99h3006L} for details. 

We now generalize these results, in particular Eq.~(\ref{eq:scalD}), to the case of a broadband spectrum of 4-velocity fluctuations extending over the range $\left[k_\text{min},\,k_\text{max}\right]$, characterized by a one-dimensional (1D) spectral index $1<q_u<2$. We note that Ref.~\cite{2013ApJ...767L..16O} has studied a similar problem, using a different formalism for turbulent shear acceleration in the subrelativistic limit. 

We therefore write
\begin{equation}
  \left\langle \updelta u^2\right\rangle = \int_{\ln k_\text{min}}\text{d}\ln k \left\langle\vert\updelta u_k\vert^2\right\rangle_k,
\end{equation}
with $\left\langle\vert\updelta u_k\vert^2\right\rangle_k \propto k^{1-q_u}$ the typical 4-velocity perturbation on scale $k^{-1}$. We also approximate: 
\begin{equation}
    \left\langle \left(\partial u\right)^2\right\rangle_k \equiv \alpha_u k^2 \left\langle \vert\updelta u_k\vert^2\right\rangle_k,
\end{equation}
with $\alpha_u$ a factor of the order of unity that depends on the properties of the turbulence. For instance, Alfv\'en waves do not contain a compressive term, while magnetosonic modes do.

If $k_\text{max}^{-1} \ll t_\text{scatt} \ll k_\text{min}^{-1}$, one can split the turbulence in a large-scale and a small-scale cascade around $t_\text{scatt}$, so that the diffusion coefficient receives two contributions (of comparable magnitude):
\begin{eqnarray}
\left. D_{pp}\right\vert_{k^{-1}\gtrsim t_\text{scatt}}& \simeq &  p^2 t_\text{scatt}\left\langle \updelta u^2\right\rangle k_\text{min}^2 \nonumber\\
&&\times\int^{\ln t_\text{scatt}^{-1}}\mathrm{ d}\ln k \left(\frac{k}{k_\text{min}}\right)^{3-q_u} \nonumber\\
&\simeq & p^2 \left(t_\text{scatt}k_\text{min}\right)^{q_u-2}\left\langle \updelta u^2\right\rangle k_\text{min},\nonumber\\
&&\nonumber\\
\left. D_{pp}\right\vert_{k^{-1}\lesssim t_\text{scatt}}& \simeq\,& p^2 t_\text{scatt}^{-1}\left\langle \updelta u^2\right\rangle\nonumber\\
&&\times \int_{\ln t_\text{scatt}^{-1}}\mathrm{ d}\ln k \left(\frac{k}{k_\text{min}}\right)^{1-q_u} \nonumber\\
&\simeq&p^2\left(t_\text{scatt}k_\text{min}\right)^{q_u-2}\left\langle \updelta u^2\right\rangle k_\text{min}.\nonumber\\
&&
\label{eq:nonres}
\end{eqnarray}
Interestingly, the total contribution scales as
\begin{equation}
    D_{pp}\propto p^{2-\epsilon} \langle\delta u^2\rangle,
\label{eq:Dppnr}
\end{equation} 
with $\epsilon$ positive but small compared to unity, because $t_\text{scatt}$ usually has a mild dependence on $p$, while $q_u-2$ is negative and small in magnitude. Using for instance the quasilinear scaling for isotropic turbulence, $t_\text{scatt} \propto \left(r_\text{g}k_\text{min}\right)^{2-q}$, we find $D_{pp} \propto p^2\left(r_\text{g}k_\text{min}\right)^{-(2-q)(2-q_u)}$.
One may furthermore note that the above result, Eq.~(\ref{eq:nonres}), departs from the naive scaling $t_\text{acc}=p^2/D_{pp}\simeq t_\text{scatt}/\langle\updelta u^2\rangle$, because the particle is now sensitive to the detailed structure of the turbulent power spectrum. Indeed, one now obtains
\begin{equation}
t_\text{acc}\approx \frac{k_\text{min}^{-1}\left(t_\text{scatt}k_\text{min}\right)^{2-q_u}}{\left\langle \updelta u^2\right\rangle}.
\end{equation}

If $t_\text{scatt} \gtrsim k_\text{min}^{-1}$, however, the above integration procedure now gives 
\begin{equation}
\left.D_{pp}\right\vert_{k_\text{min}^{-1} \lesssim t_\text{scatt}} \simeq p^2 \frac{\left\langle \updelta u^2\right\rangle}{t_\text{scatt}}.
\label{eq:nonres2}
\end{equation}
In particular, for $r_\text{g}k_\text{min}\gg1$, we have $t_\text{scatt}\gg k_\text{min}^{-1}$ and $t_\text{scatt}\propto r_\text{g}^2$; hence, we recover $D_{pp}\propto p^0$.

\subsection{Non-MHD electric fields}
Ideal Ohm's law $\boldsymbol{E} =-\boldsymbol{\beta}\times\boldsymbol{B}$ prevents the existence of electric fields aligned with the local magnetic field and guarantees that there exists at every point a reference frame in which the electric field can be screened out (the local plasma rest frame). 

In the presence of a parallel electric field, acceleration could proceed at a much faster rate, unhindered by the magnetic field. On the largest physical scales, however, it is believed that the ideal Ohm's law provides a satisfactory approximation and our simulations have implemented this constraint. Ideal MHD rather breaks down on small length scales, generating parallel electric fields in reconnecting current sheets~\cite{17Loureiro,*17Mallet}, or, more generally, because of inertial and kinetic effects, e.g., Ref.~\cite{2010A&A...519A.114B}. Of course, if the typical current sheet width is $l_\perp$, then particles with gyroradius $r_\text{g} \gg l_\perp$ stream through the sheet with small deflection and energy gain; hence, such reconnection processes govern the small-scale physics of dissipation, heating, and injection, but not acceleration to high energies.

Our aim, here, is to quantify the influence of these small-scale violations of Ohm's law on the acceleration process. We first characterize the statistics of these electric field fluctuations through a power spectrum, assuming that they extend on spatial scales greater than $k_\text{max}^{-1}$ but peak on $k_\text{max}^{-1}$ with, correspondingly, a 1D index $q_E<1$ ($q_E$ can possibly take negative values). In Appendix~\ref{sec:appE}, we calculate the relevant index from the corrections to Ohm's law for Alfv\'en, fast and slow wave turbulence respectively, for pair and electron-proton plasmas. 

The general problem can be brought back to the one-dimensional analog that describes the evolution of the particle momentum through
\begin{equation}
    \frac{\text{d}p}{\text{d}t} = q \mu \updelta E_\parallel,
    \label{eq:1dE1}
\end{equation}
where $\updelta E_\parallel$ is understood here as a random field and $\mu$ represents the effective (random) velocity of the particle, characterized by a time correlation function of step $t_\text{scatt}$. Note that this scattering timescale may differ from that introduced in the previous sections, as it may be affected by the small-scale electric field itself. 

Equation~(\ref{eq:1dE1}) describes a diffusion process, because over a time interval $\Delta t$, $\langle \Delta p\rangle=0$ but $\langle\Delta p^2\rangle\neq0$. 

For explicit calculations, we adopt for the velocity correlation function
\begin{equation}
    \langle\mu(t)\mu(0)\rangle=\mu(0)^2\exp\!\left[-\frac{\uppi}{4}\frac{t^2}{t_\text{scatt}^2}\right]\!.
    \label{eq:corrmu}
\end{equation}
As it should, the scattering timescale verifies $t_\text{scatt}=\int_0^{+\infty}\text{d}t\,\langle\mu(t)\mu(0)\rangle/\mu(0)^2$.

In the spirit of the previous sections, we decompose the electric field fluctuations over scales $k^{-1}$  as a power-law spectrum and we further assume that at each such scale, the coherence length of the corresponding mode is also $k^{-1}$. This generalized decomposition in wave packets mimics the decomposition of the damped modes of the Alfv\'en and slow modes.  Hence, we use
\begin{equation}
\langle\updelta E_\parallel(t_1)\updelta E_\parallel(t_2)\rangle=\alpha_E \langle\updelta E_\parallel^2\rangle\! \int \mathrm{d}k\,k^{-q_E}\exp\!\left[-k \vert t_1-t_2\vert\right]\!.
\label{eq:corrE}
\end{equation}
The exponential term involves a correlation time $k^{-1}$, which can be seen as the turnover timescale of the parallel electric field structure in the turbulence.  Note that, using linear or squared expressions in the exponentials, either for Eq.~(\ref{eq:corrmu}) or for Eq.~(\ref{eq:corrE}), would modify our final results by a factor of the order of unity only. The prefactor $\alpha_E=\vert1-q_E\vert\,k_\text{max}^{q_E-1}$ provides the correct normalization to the equal-time (equal-position) amplitude $\langle\updelta E_\parallel^2\rangle$.

We thus derive the momentum diffusion coefficient as
\begin{align}
    \langle\Delta p^2\rangle& =
    2\Delta t\,q^2\langle\updelta E_\parallel^2\rangle\alpha_E\mu(0)^2\,t_\text{scatt}\nonumber\\
    &\quad\times \int \text{d}k\,k^{1-q_E}\,\e^{k^2 t_\text{scatt}^2 /\uppi}\,\text{Erfc}\!\left[\frac{k t_\text{scatt}}{\sqrt{\uppi}}\right]\!.
    \label{eq:diffE1}
\end{align}
The complementary error function is here defined as $\text{Erfc}(x)=(2/\sqrt{\uppi})\int_x^{+\infty}\text{d}t\,\e^{-t^2}$. Integrating this contribution over $k$ then gives the approximate diffusion coefficient $D_{pp}^{\updelta E_\parallel}=\langle\Delta p^2\rangle/2\Delta t$:
\begin{align}
    D_{pp}^{\updelta E_\parallel}& \approx q^2\langle\updelta E_\parallel^2\rangle t_\text{scatt} &  \left(t_\text{scatt}\ll k_\text{max}^{-1}\right)\,\nonumber\\
    & \nonumber\\
    &\approx q^2\langle\updelta E_\parallel^2\rangle \frac{\left( t_\text{scatt}k_\text{max}\right)^{q_E}}{k_\text{max}} & \left(t_\text{scatt}\gg k_\text{max}^{-1},\, 0<q_E\right)\,\nonumber\\
    & \nonumber\\
    &\approx q^2\langle\updelta E_\parallel^2\rangle k_\text{max}^{-1} & \left(t_\text{scatt}\gg k_\text{max}^{-1},\, q_E<0\right)\!.\nonumber\\
    \label{eq:diffE2}
\end{align}
This diffusion coefficient thus increases up as the scattering timescale until the particle starts to see these fluctuations as small scales, meaning $t_\text{scatt}\gtrsim k_\text{max}^{-1}$, at which point it may either level off, if $q_E<0$ (corresponding to a spectrum that is sharply peaked at $k_\text{max}$), or increase as $t_\text{scatt}^{q_E}$, if the long-wavelength fluctuations at $k^{-1}>t_\text{scatt}$ retain enough power, meaning $0<q_E<1$. We note that, in Eq.~(\ref{eq:diffE2}), the first approximation indeed describes the diffusion coefficient of a particle changing direction every $t_\text{scatt}$ while traveling in a roughly coherent electric field, while the third approximation describes how a particle gains energy by traveling ballistically over many incoherent patches of parallel electric field of typical scale $k_\text{max}^{-1}$. In the intermediate limit, the particle feels the structure of the power spectrum of electric field fluctuations. For reference, we derive in Appendix~\ref{sec:appE}, $q_E\simeq -7/3,\,-5/3,\,-1$ respectively, for isotropic fast mode, Goldreich-Sridhar slow mode, and Goldreich-Sridhar Alfv\'en mode turbulences, from the deviations to the ideal Ohm's law in a pair plasma. 

Provided $t_\text{scatt}^{q_E}$ roughly scales less fast with energy than a square law, acceleration by the large-scale MHD turbulence, as described in the previous sections, eventually takes over at some energy which may be easily calculated from the above expressions, once $t_\text{scatt}(p)$, $q_E$, and $\langle \updelta E_\parallel^2\rangle$ are specified. To ease the comparison, we note that the units in which we express $D_{pp}$ in Figs.~\ref{plot:F_Dp}, \ref{plot:SGS_Dp}, and \ref{plot:AGS_Dp} are $q^2B_0^2/(k_\text{min}c)$, while those for $D_{pp}^{\updelta E_\parallel}$ are  $q^2\langle\updelta E_\parallel^2\rangle/(k_\text{max}c)$.

\section{Discussion}
\label{sec:disc}

The physics of particle acceleration in a turbulent setting is governed by a variety of effects, depending on whether the cascade can be described as isotropic or not, whether it can be approximated as linear waves or not, whether one assumes ideal MHD to hold or not. In this section, we recap and bring together the results obtained in the previous sections and compare them to the results of recent first-principles extensive numerical simulations of turbulent acceleration.

\subsection{Comparison between modes and general results}
In an actual turbulent setting, one expects a mixed contribution from various modes. As mentioned earlier, numerical MHD simulations of subrelativistic turbulence generally point to the dominance of Alfv\'en and slow modes, with a minor contribution from fast magnetosonic modes~\cite{03Cho,10Kowal,17Andre}. This result, however, appears to depend on how the turbulence is driven at the outer scale~\cite{2019arXiv190701853M}, and the phenomenology may differ in the  relativistic regime~\cite{16Takamoto}. We can nevertheless combine our various results, adopting the notations $\eta_\text{F}$, $\eta_\text{A}$, and $\eta_\text{S}$ for the respective contributions of the various modes to the total magnetic energy density.

Our numerical simulations for acceleration in fast, Alfv\'en and slow mode turbulences provide the following scalings in the inertial range $r_\text{g}k_\text{min}<1$,
\begin{align}
   &D^\text{F}_{pp} \simeq 0.5\frac{\eta_\text{F}}{1-\eta} \left(r_\text{g}k_\text{min}\right)^{q_F}\!\left(\beta_\text{A}/0.9\right)^2\!, \nonumber\\
   &D^\text{A}_{pp}\simeq 0.07 \frac{\eta_\text{A}}{1-\eta}\left(r_\text{g}k_\text{min}\right)^2\left[1-2.7\ln \! \left(r_\text{g}k_\text{min}\right)\right]\!\left(\beta_\text{A}/0.9\right)^3\!,\nonumber\\
   &D^\text{S}_{pp} \simeq 0.01 \frac{\eta_\text{S}}{1-\eta} \left(r_\text{g}k_\text{min}\right)^2 \left[1-1.8\ln\! \left(r_\text{g}k_\text{min}\right)\right]\!\left(\beta_\text{S}/0.5\right)^2\!, \label{eq:scalDall}
\end{align}
where all diffusion coefficients are here written in units of $m^2 \Omega_0^2/k_\text{min}$, with $\Omega_0=eB_0/m$ the cyclotron frequency. Note that $m^2 \Omega_0^2/k_\text{min}=p_\text{conf}^2\,k_\text{min}$, with $p_\text{conf}$ the confinement momentum such that $r_\text{g}\!\left(p_\text{conf}\right)k_\text{min}=1$.
We rely here on the (more realistic) picture of damped modes for Alfv\'en and slow modes, and we have used $q_\text{F}=5/3$, $q_\text{S}=q_\text{A}=7/3$ in our simulations. The wave number $k_\text{min}$ is related to the outer scale of the turbulence through $k_\text{min}=2\pi/L_\text{max}$. Note that the effective slow mode phase speed $\beta_\text{S}$ cannot be realistically larger than about $0.5$ because it is bounded by the sound speed, which is itself bounded by $1/\sqrt{3}\simeq0.58$ in a relativistically hot plasma. Note also the scaling $D_{pp}\propto \beta_\text{A}^3$ for Alfv\'en modes, to be contrasted with the naive square law. Here, one extra power comes from $\Im\omega\propto\Re\omega$, which controls the scattering through gyroresonance broadening. We have omitted this contribution for slow modes, since our analysis has revealed that pitch-angle randomization provides an equally strong source of scattering. Finally, in the region $r_\text{g}k_\text{min}\gg1$, these scalings are each continued into a constant value $D_{pp}\propto \eta\,m^2k_\text{min}$.

The acceleration timescale is conveniently written as $t_\text{acc}=\left(r_\text{g}k_\text{min}\right)^2 \left(m^2\Omega_0^2k_\text{min}^{-1}/D_{pp}\right)k_\text{min}^{-1}$. We thus derive the following scalings:
\begin{align}
   &t^\text{F}_\text{acc} \simeq 2.1\frac{1-\eta}{\eta_\text{F}}\,\left(r_\text{g}k_\text{min}\right)^{2-q_F}\!\left(\beta_\text{A}/0.9\right)^{-2}k_\text{min}^{-1},  \nonumber\\
   &t^\text{A}_\text{acc} \simeq 14 \frac{1-\eta}{\eta_\text{A}} \left[1-2.7\ln\!\left(r_\text{g}k_\text{min}\right)\right]^{-1}\!\left(\beta_\text{A}/0.9\right)^{-3} k_\text{min}^{-1}, \nonumber\\
   &t^\text{S}_\text{acc}\simeq 100 \frac{1-\eta}{\eta_\text{S}} \left[1-1.8\ln\!\left(r_\text{g}k_\text{min}\right)\right]^{-1}\!\left(\beta_\text{S}/0.5\right)^{-2} k_\text{min}^{-1}. \nonumber\\
    \label{eq:scaltaccall}
\end{align}
If one rather defines the acceleration timescale as the timescale for the mean energy to double in the Monte Carlo simulations, the prefactors in Eq.~(\ref{eq:scaltaccall}) are found to be instead approximately equal to $2$, $4$, and $30$ for fast modes, damped Alfv\'en modes, and damped slow modes, respectively.

In Sec.~\ref{sec:nra}, we have also discussed the physics of acceleration of particles in a more generic turbulent setting, which is not described as a sum of linear waves but whose velocity field is decomposed into a sum of compressive, shearing, vortical, and accelerating motions. Although the scattering timescale cannot be predicted on general grounds in such a situation, we have found that, for $t_\text{scatt}\,\lesssim\,k_\text{min}^{-1}$, $D_{pp}\sim p^2\,\langle\delta u^2\rangle\,k_\text{min}\, \left(t_\text{scatt}k_\text{min}\right)^{2-q_u}$ in terms of the power spectrum index of the 4-velocity fluctuations, $q_u$. We have argued there that, since $t_\text{scatt}$ is generically a mild (increasing) function of $p$, and since $2-q_u$ is small compared to unity, $D_{pp}$ can be written in the general form 
\begin{equation}
    D_{pp} \sim p^2 \langle\delta u^2\rangle k_\text{min},
    \label{eq:scalDnr}
\end{equation}
i.e., $D_{pp}\sim \langle\delta u^2\rangle\left(r_\text{g}k_\text{min}\right)^2$ in units of  $m^2\Omega_0^2/k_\text{min}$. The above estimate is valid up to a correction of the order $\left(r_\text{g}k_\text{min}\right)^{-\epsilon}$, with $\epsilon$ significantly smaller than unity. Note that the amplitude of the turbulence, $\eta$, is included here in the 4-velocity fluctuation amplitude $\langle\delta u^2\rangle$.

\begin{figure}
\center \includegraphics[scale=1.]{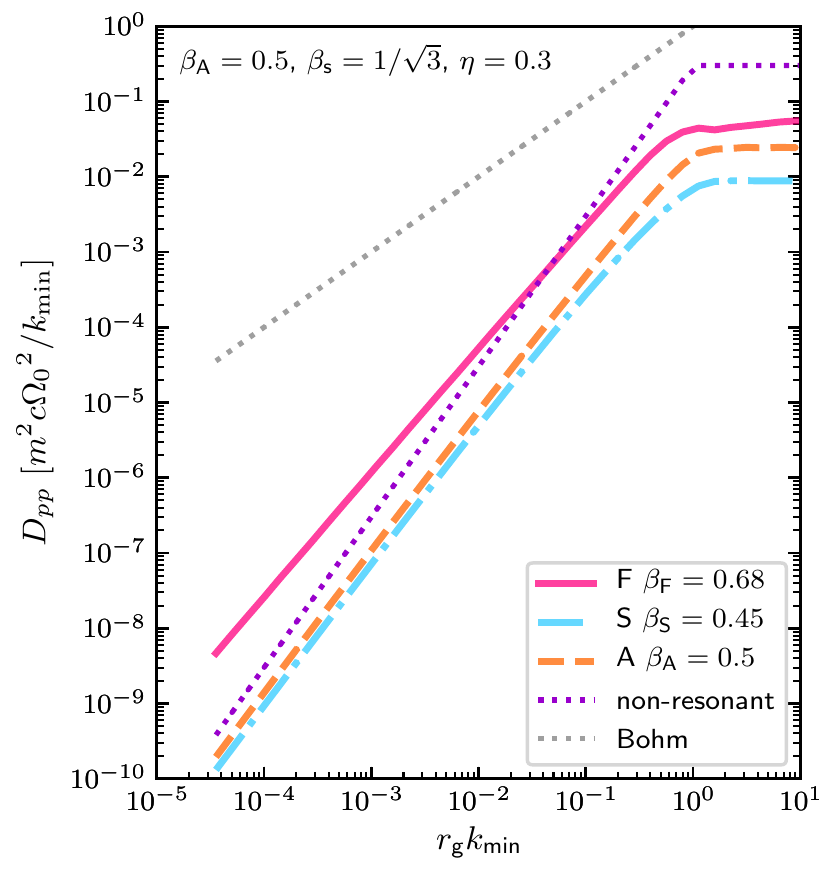}
\caption{Comparison of the predicted momentum diffusion coefficients for various turbulence modes for a relativistically hot plasma ($\beta_\text{s}\simeq0.58$) with $\beta_\text{A}=0.5$ and $\eta=0.3$ (corresponding to $\updelta B/B\simeq 0.7$). For fast modes, this corresponds to $\beta_\text{F}\approx 0.68$, while for slow modes, to $\beta_\text{S}\approx0.45$. In the case of Alfv\'en and slow modes, we considered damped modes with $\vert\Im\omega\vert=\vert\Re\omega\vert$ (i.e., $\gamma_\text {d}=1$). The purple dotted line shows the theoretical prediction $D_\text{pp}\sim \eta\, p^2 k_\text{min}$, which corresponds to nonresonant acceleration (Sec.~\ref{sec:nra} and see the text for details), and $D_{pp}\simeq \eta  m^2\Omega_0^2/k_\text{min}$ at $r_\text{g}k_\text{min}\gg1$. The gray dotted line indicates the Bohm scaling for comparison, $D_\text{pp}\sim  r_\text{g}k_\text{min}\,m^2\Omega_0^2/k_\text{min}$.
\label{plot:Dpp_comp}
}
\end{figure}
\begin{figure}
\center \includegraphics[scale=1.]{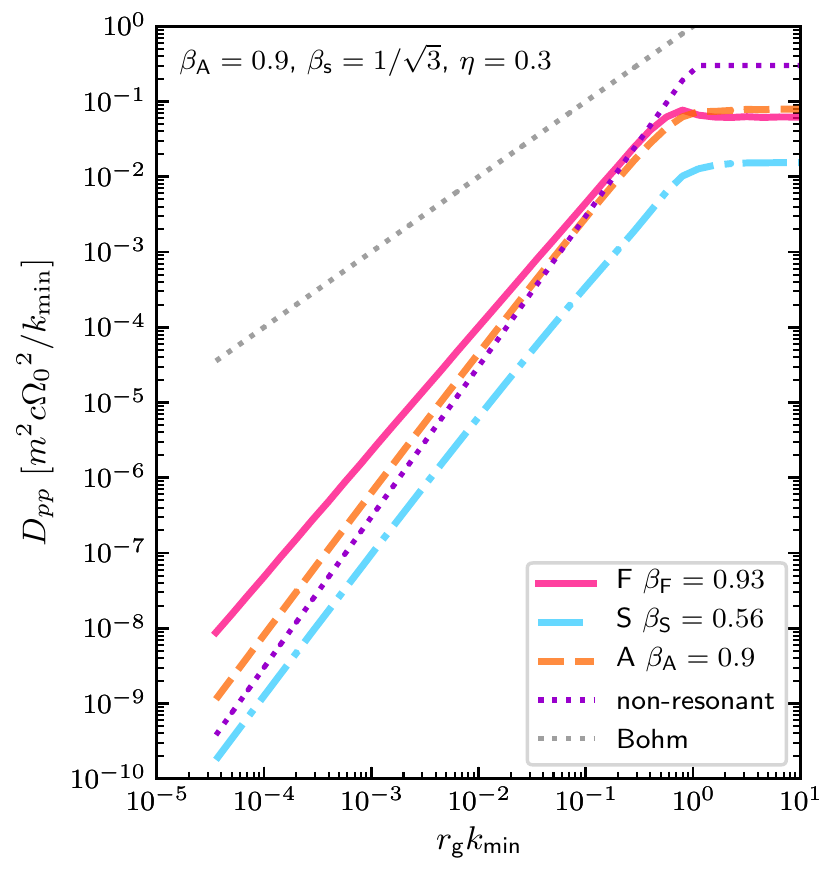}
\caption{Same as Fig.~\ref{plot:Dpp_comp}, with this time $\beta_\text{A}=0.9$, so that $\beta_\text{F}\approx 0.93$, while $\beta_\text{S}\approx0.56$.} 
\label{plot:Dpp_comp2}
\end{figure}

These different contributions are brought together in Figs.~\ref{plot:Dpp_comp} and \ref{plot:Dpp_comp2} for the case of a relativistically hot plasma with magnetization of order unity for $\eta_\text{F,S,A}=\eta=0.3$. 
Fig.~\ref{plot:Dpp_comp2} reveals that, for $\beta_\text{A}\simeq 1$, the contributions of fast and Alfv\'en  modes are roughly comparable at large rigidities, $r_\text{g}k_\text{min}\gtrsim 0.1$,\footnote{Especially given that the predictions for fast modes displayed here include resonance broadening due to the pitch-angle uncertainty which tend to slightly overestimate the diffusion coefficients at large rigidities.} but that the fast mode contribution dominates at lower rigidities due to its slightly softer dependence on $r_\text{g}$. Recall however that this dependence scales directly with $q_\text{F}$, and if $q_\text{F}\sim2$, the fast mode would no longer dominate over Alfv\'en modes.

In a relativistically hot, magnetized plasma, the phase speed of the slow modes can reach mildly relativistic values; hence, slow mode acceleration becomes truly efficient as well. In particular, we find that, at equal (relativistic) phase velocities, the diffusion coefficient in slow modes is only a factor of a few below that of Alfv\'en modes (Fig.~\ref{plot:Dpp_comp}). This hierarchy differs strongly from what is observed in the subrelativistic regime, namely a strong dominance of fast modes over Alfv\'en modes, with a negligible contribution from slow modes. This arises as a combination of several effects, notably the partial disappearance of transit-time damping for fast modes, the partial restoration of TTD for slow modes due to pitch-angle randomization, and the partial restoration of gyroresonances for Alfv\'en modes due to the finite lifetime of the modes.

We also observe that our predictions for nonresonant acceleration, i.e., acceleration in a turbulence whose spectrum is composed of structures rather than linear waves, also match the above resonant values, at least in the range of momenta considered, at the upper end of the inertial range.

\begin{figure}
\center \includegraphics[scale=1.]{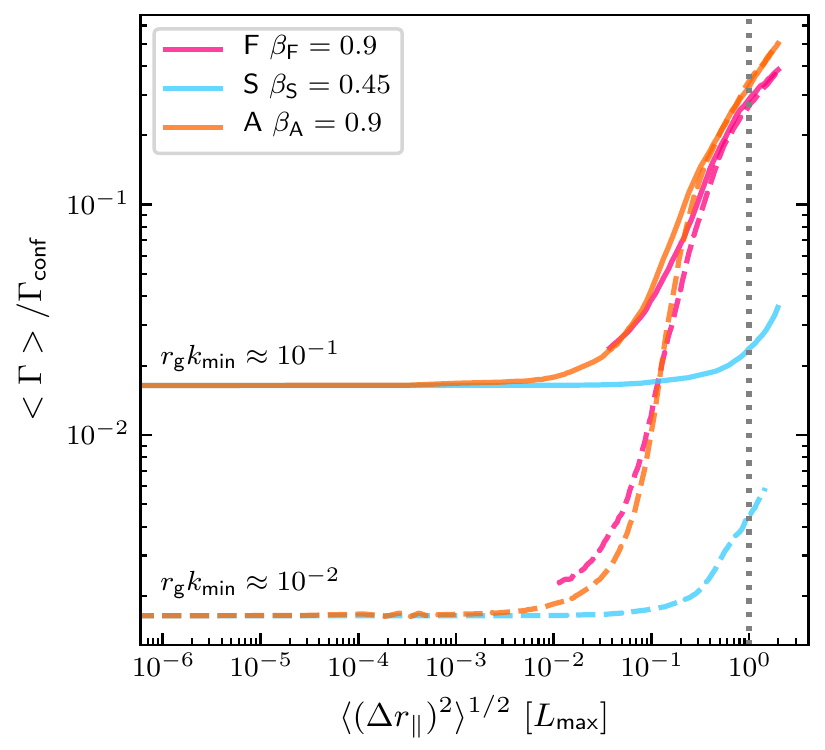}
\caption{Evolution of the energy of particles relative to the confinement energy at which $r_\text{g}=2\pi/ k_\text{min}$ vs the distance traveled along the parallel direction in units of $2\pi/k_\text{min}=L_\text{max}$, for two different injection rigidities, namely $r_\text{g}k_\text{min}\approx0.01$ and $0.1$, for simulations of fast modes, damped slow modes and damped Alfv\'en modes, as indicated in the legend. The data have been rescaled to $\eta=0.3$, corresponding to $\updelta B/B\approx 0.7$. The data for fast modes have also been rescaled to remove the initial jump in energy.}
\label{plot:tacc_comp2}
\end{figure}

An interesting outcome of our test-particle Monte Carlo simulations is to reveal that, as the particle gains energy through its stochastic interactions, it undergoes superdiffusive spatial transport along the parallel direction. This has consequences for the maximal energy of acceleration, in particular whether a given particle can reach the confinement energy $r_\text{g}\sim R_\text{s}$, with $R_\text{s}$ the size of the source, or not. Assuming $R_\text{s}\sim L_\text{max}$, we plot in Fig.~\ref{plot:tacc_comp2} the evolution of the particle energy in units of the confinement energy as a function of the distance traveled in the parallel direction. 

\subsection{Comparison to recent numerical results}
Recently, several groups have provided the first PIC numerical simulations of relativistic turbulence~\cite{17Zhdankin,*2018MNRAS.474.2514Z,*2018ApJ...867L..18Z,*18Zhdankin,18Comisso,19Wong,2019arXiv190901420C}. The merit of such simulations is to provide a self-consistent description of the nonlinear interaction between the thermal plasma, the nonthermal population and the electromagnetic fields. The main disadvantages of such simulations are, of course, their high numerical cost and their need to resolve (finely enough) the skin depth scale of the plasma. The numerical cost has to be balanced against the number of particles involved, fewer particles per cell meaning a higher numerical noise. Resolving the skin depth scale allows to properly model the small-scale dissipative physics, which is missing in MHD simulations, but it also means that the maximal energy at acceleration is limited by the dynamic range of the simulation, just as the acceleration mechanism may be affected by small-scale physical effects which would be absent on the large scales of astrophysical sources.

Nevertheless, the dynamical range of the recent simulations shown in Refs.~\cite{18Comisso,19Wong,2019arXiv190901420C} is so large that one may expect such effects to be absent. Such simulations thus provide a perfect experimental benchmark for our results. These simulations find that acceleration proceeds in two stages. The first is rapid and likely associated to particle energization in current sheets, through, e.g., reconnection-type acceleration, as discussed in these studies. The second is slower and presumably proceeds through stochastic acceleration in a turbulence that mostly obeys the ideal Ohm's law. In particular, Ref.~\cite{19Wong} derives a diffusion coefficient $D_{pp}\propto p^{2/3}$ in the first stage, but $D_{pp}\propto p^2$ in the second; see also Ref.~\cite{2019arXiv190901420C}. Our discussion of the previous sections agrees well with such a picture. If the small-scale physics can be described in terms of acceleration in parallel electric fields whose power lies on scale $k_\text{max}$, we have found that $D_{pp} \propto t_\text{scatt}$ for $t_\text{scatt}\ll k_\text{max}^{-1}$, while  $D_{pp}\propto t_\text{scatt}^{q_E}$ for $t_\text{scatt}\gg k_\text{max}^{-1}$ if $0<q_E<1$. Consider, for instance, the latter possibility; the scaling of Ref.~\cite{19Wong} is then  recovered if $t_\text{scatt}\propto p^{2/(3q_E)}$. Alternatively, if $t_\text{scatt}\propto p^2$, describing the limit in which the particle at such energies is sensitive to the electric field structures on scales $k_\text{max}$ but not to the large-scale MHD turbulence, then $q_E=1/3$ would explain the observed scaling.  A detailed study of the statistics of the small-scale parallel electric fields and a careful follow-up of the particle momentum in the corresponding PIC simulations would allow to test such hypotheses.

Regarding the acceleration on larger scales, the main result of Refs.~\cite{19Wong,2019arXiv190901420C} is $t_\text{acc}\sim L_\text{max}/\langle\delta u^2\rangle$. This agrees well with the scalings given previously, since it corresponds to $D_{pp} \sim 0.16\, \eta\beta_\text{m}^2\left(r_\text{g}k_\text{min}\right)^2$ in units of $m^2\Omega_0^2/k_\text{min}$, with $\beta_\text{m}$ the relevant mode velocity, for direct comparison to Eqs.~(\ref{eq:scalDall}) and (\ref{eq:scalDnr}). In particular, it broadly agrees with acceleration in Alfv\'en and/or slow mode turbulence, accounting for wave decay, or with nonresonant acceleration in a generic turbulence. In order to better understand which of these description better applies, one would need to perform a mode decomposition of the turbulence as in Ref.~\cite{03Cho}, and to track the scaling of $t_\text{scatt}$ with energy in these PIC simulations.

\section{Conclusions}\label{sec:summ}
This work has studied the physics of particle acceleration in relativistic turbulence, where relativistic means typical 3-velocity fluctuations at the outer scale $\langle\delta \beta^2\rangle^{1/2}\sim 1$. For an MHD turbulence described in terms of eigenmodes, fast and slow magnetosonic, and Alfv\'en, this corresponds to a magnetized, relativistically hot plasma for which $\beta_\text{A}$ is close to unity (and $\beta_\text{s}=1/\sqrt{3}$); hence, $\sigma\gtrsim0.1$ in terms of magnetization $\sigma=\beta_\text{A}^2\big/\left(1-\beta_\text{A}^2\right)$.

We have provided detailed analytical estimates of the pitch-angle scattering ($D_{\mu\mu}$) and momentum ($D_{pp}$) diffusion coefficient in quasilinear, and extended quasilinear theory for acceleration in isotropic fast wave turbulence, as well as in Goldreich-Sridhar-like anisotropic Alfv\'en and slow wave turbulence. In turn, this allows to provide analytical scalings for the scattering ($t_\text{scatt}$) and acceleration ($t_\text{acc}$) timescales. We have notably included the effect of a finite mode lifetime, as expected for strong turbulence, and discussed in some detail the resonance broadening associated to the partial randomization of the pitch angle of the particle. We have compared these predictions to dedicated Monte Carlo simulations of test-particle acceleration in synthetic wave turbulence. We have paid attention to the notion of local anisotropy, which is inherent to the Goldreich-Sridhar phenomenology of anisotropic turbulence. To do so, we conduct these anisotropic simulations with a low enough turbulence amplitude to guarantee that the eddies are effectively aligned with respect to the mean field on the scale of the particle gyroradius.
Our Monte Carlo simulations naturally account for the partial randomization of the pitch angle and they include, where necessary, the finite lifetime of turbulent modes. The satisfactory agreement that we reach between those analytical scalings and the numerical simulations suggests that our analytical predictions capture the salient effects of particle acceleration in a prescribed wave turbulence.

We notably observe the following effects in a relativistic turbulent setting:
\begin{enumerate}[label={(\arabic*)}]
    \item For an isotropic cascade of fast magnetosonic modes the transit-time damping contribution is progressively reduced as $\beta_\text{F}$ increases, because the resonant value of the longitudinal phase velocity of the wave is then forced into the superluminal regime. Even though resonance broadening effects preserve part of the transit-time damping contribution, most of the acceleration occurs through gyroresonant interactions. The momentum diffusion coefficient accordingly scales as $D_{pp}^\text{F}\sim \beta_\text{F}^2\eta\, p^2 k_\text{min}\!\left(r_\text{g}k_\text{min}\right)^{q-2}$ for $r_\text{g}k_\text{min}\lesssim1$, with $q$ the 1D index of the turbulence spectrum and $k_\text{min}^{-1}=L_\text{max}/(2\uppi)$ in terms of the outer scale $L_\text{max}$.
    
    \item For an anisotropic, Goldreich-Sridhar-like cascade of slow mode waves, the transit-time damping contribution provides the dominant  contribution to scattering and acceleration, thanks to the resonance broadening implied by the partial randomization of the pitch-angle. Gyroresonant-like interactions are negligible. The momentum diffusion coefficient accordingly scales as $D_{pp}^\text{S}\sim \beta_\text{S}^2\eta \,p^2 k_\text{min}$ for $r_\text{g}k_\text{min}\lesssim1$, up to a logarithmic correction. 
    
    \item For an anisotropic, Goldreich-Sridhar-like cascade of Alfv\'en waves, the resonance broadening associated to the finite lifetime of the modes restores gyroresonant-like interactions, that would otherwise be inefficient. The partial randomization of the pitch angle does not play any significant role. The momentum diffusion coefficient scales according to $D_{pp}^\text{A}\sim \beta_\text{A}^3\eta \,p^2 k_\text{min}$ for $r_\text{g}k_\text{min}\lesssim1$, up to a logarithmic correction, as for slow modes. The extra power of $\beta_\text{A}$ results from resonance broadening.
    
    \item At large rigidity, $r_\text{g}k_\text{min}\gtrsim1$, the particle effectively sees small-scale turbulence, hence $D_{pp}\sim \beta^2\eta\, m^2\Omega_0^2k_\text{min}^{-1}\propto p^0$ for all three types of modes, with $\beta$ the relevant phase velocity.
    
    \item Overall, we find that, in a relativistic setting, if all three types of modes share a similar fraction of the turbulent energy, they give roughly comparable contributions to the acceleration of particles, for momenta not far below the top of the inertial range where $r_\text{g}\sim k_\text{min}^{-1}$. At lower rigidities, $r_\text{g}\ll k_\text{min}^{-1}$, the contribution of fast modes becomes dominant, because of its softer scaling with $r_\text{g}$.

    \item We have also provided general arguments concerning the acceleration of particles in a turbulence that cannot be described as a bath of linear waves, but rather as a combination of compressive, shearing, vortical and accelerating fluid motions obeying the ideal Ohm's law. We have shown, notably, that if the scattering timescale of particles is such that $t_\text{scatt}\lesssim k_\text{min}^{-1}$ (as for all three types of modes above at $r_\text{g} \ll k_\text{min}^{-1}$), then the diffusion coefficient $D_{pp}\sim \langle\delta u^2\rangle\,p^2 k_\text{min}$ up to a correction factor $\left(t_\text{scatt}k_\text{min}\right)^{q_u-2}$ that depends weakly on energy. Here, $q_u$ represents the index of the 4-velocity 1D turbulent spectrum, and $\langle\delta u^2\rangle$ represents its amplitude. 
    
    \item We have also discussed the possible contributions of violations of ideal Ohm's law, showing that they peak on the small scales $\sim k_\text{max}^{-1}$ of the turbulent cascade, and we have characterized their magnitude. We notably find that, if the 1D power spectrum index $q_E$ of the parallel electric field component verifies $0<q_E<1$ and $t_\text{scatt}\gg k_\text{max}^{-1}$, then such small-scale effects provide a contribution $D_{pp}\sim q^2\langle\updelta E_\parallel^2\rangle k_\text{max}^{-1}\!\left( t_\text{scatt}k_\text{max}\right)^{q_E}$ which may dominate the acceleration at very small rigidities.

    \item We have compared our results to recent {\it ab initio} simulations of turbulence using kinetic particle-in-cell simulations and shown that the above generally agree with the observed results. 
\end{enumerate}
    
Finally, we provide ready-to-use analytical scalings for applications to high-energy astrophysical phenomenology in Sec.~\ref{sec:disc}. 

\section*{Acknowledgments}
We thank the anonymous referees for useful suggestions.

This work has been financially supported by the ANR-14-CE33-0019 MACH project. C.D. and F.C. acknowledge the financial support from the UnivEarthS Labex program  of Sorbonne Paris Cit\'e (ANR-10-LABX-0023 and ANR-11-IDEX-0005-02). C.D. also gratefully acknowledges the support of the Japan Society for the Promotion of Science (JSPS International Research Fellowship at the Astrophysical Big Bang Laboratory, RIKEN). M.L. acknowledges support by the Emergence-2019 SU program and by the National Science Foundation under Grant No.~NSF PHY-1748958. This work was granted access to HPC resources of Institut du d\'eveloppement et des ressources en informatique scientifique (IDRIS) and Centre Informatique National de l\textsc{\char13}Enseignement Sup\'erieur (CINES) under the allocation  A0040410126 made by Grand Equipement National de Calcul Intensif (GENCI). 

\appendix

\section{RESONANCE BROADENING\\ IN QUASILINEAR THEORY}\label{sec:Appbrd}
The theoretical estimates of the diffusion coefficients, Eqs.~(\ref{eq:QLFMSDmumu}) and (\ref{eq:QLAGSDmumu}) for pitch-angle diffusion, Eqs.~(\ref{eq:QLFMSDpp}) and
(\ref{eq:QLAGSSDpp}) for momentum diffusion, involve resonance functions  $\mathcal R_{\boldsymbol{k}}$ that characterize the interaction between particles and waves. At a formal level, this resonance function derives from the time-integrated Fourier transform of the propagator that connects the position of the particle at different times in the turbulent bath, e.g., Ref.~\cite{1966PhFl....9.1773D}. Accounting for gyromotion around the background magnetic field, this resonance function is expressed as
\begin{equation}
    \mathcal R_{\boldsymbol{k}} = \frac{1}{2\uppi}\int_{-\infty}^{+\infty}\text{d}\tau\, \e^{\i(k_\parallel\mu-\omega+n\Omega)\tau}.
    \label{eq:prop}
\end{equation}
In standard quasilinear theory, with $\omega\in\mathbb{R}$ and $\mu$, the initial  pitch-angle cosine of the particle, this resonance becomes a Dirac-function generating the transit-time damping and the infinite harmonic series of gyrosynchrotron resonances, as explained in the text. 

It has long been appreciated, however, that these idealized resonances are actually broadened to some degree by various physical effects. In our case, the two major causes of broadening are the finite lifetime of the turbulent modes and the partial randomization of the pitch angle of the particle. For the sake of the argument, we treat each case separately.

If the linear eigenmodes are assigned a finite lifetime, the mode frequency can be written  $\omega=\Re\omega-\i\gamma_\text{d}\vert\Re\omega\vert$, with $\gamma_\text{d}>0$. The resonance function then takes on a Breit-Wigner form,
\begin{equation}
    \mathcal R_k = \frac{1}{\uppi}\frac{\gamma_\text{d}\vert\Re\omega\vert}{\left(k\mu_k\mu - \Re\omega + n\Omega\right)^2 + \gamma_\text{d}^2\Re\omega^2},
    \label{eq:brdom}
\end{equation}
whose finite width is directly governed by $\gamma_\text{d}\vert\Re\omega\vert$.

The presence of net magnetic field fluctuations implies that the pitch angle of the particle is modulated at all times by a random quantity, which scales as some power of the fluctuation. This consequently broadens the resonance of particles with waves and permits, in particular, transit-time damping to occur in slow mode turbulence over a broad range of particle pitch angles; see Sec.~\ref{sec:S}. This effect has been introduced in the subrelativistic regime in a number of studies, starting with Ref.~\cite{1975RvGSP..13..547V}; see, e.g., Refs.~\cite{18Xu,19Teraki} for recent implementations. Here, we provide a detailed discussion of this effect in the relativistic limit, and emphasize the differences with respect to these previous studies.

If the pitch-angle cosine $\mu$ of the particle becomes a random quantity, the ballistic propagator $\exp\!\left[\i \!\left(k_\parallel\mu- \omega+n\Omega\right)\!\Delta t\right]$ can be approximated, to second-order in the cumulant expansion of $\mu$, by
\begin{equation}
    \begin{split}
    \left\langle \e^{\i\left(k_\parallel\mu-\omega+n\Omega\right)\Delta t}\right\rangle \simeq & \exp\!\biggl[\i\!\left(k_\parallel\langle\mu\rangle-\omega+n\Omega\right)\!\Delta t\nonumber\\
    &\qquad -\frac{1}{2}k_\parallel^2\langle\Delta\mu^2\rangle \Delta t^2\biggr]\,.
    \end{split}
    \label{eq:prop1}
\end{equation}
Consequently, the resonance function becomes
\begin{equation}
    \mathcal R_k = \frac{1}{\left(2\uppi k_\parallel^2\langle\Delta\mu^2\rangle\right)^{1/2}}
    \exp\!\left[-\frac{\left(k_\parallel\langle\mu\rangle-\omega+n\Omega\right)^2}
    {2k_\parallel^2\langle\Delta\mu^2\rangle}\right]\!,
    \label{eq:brdmu}
\end{equation}
and it is entirely characterized by $\langle\mu\rangle$ and $\langle\Delta\mu^2\rangle$.

For strongly magnetized particles, meaning $r_\text{g}k_\text{min}\ll1$, these moments can be estimated using the conservation laws of the first adiabatic invariant and of the energy, which guarantee that $\left(1-\overline{\mu}^2\right)/B$ is a conserved quantity along the trajectory. In this expression, $\overline{\mu}$ represents the pitch-angle cosine as measured relative to the direction of the total magnetic field $\boldsymbol{B}$. It differs from the standard pitch-angle cosine $\mu$, which is defined relative to the direction of the unperturbed field, by an amount $\updelta B_\perp^2/2B_0^2\sim \eta/2(1-\eta)$.

Let us first consider the case $\eta\ll1$, as used in Secs.~\ref{sec:S} and \ref{sec:A} to build an effective model of the anisotropic Goldreich-Sridhar phenomenology, so that $\overline\mu\simeq\mu$. Then, as the particle travels from one coherence cell of the turbulence, on scale $k_\text{min}^{-1}$, to another, its pitch-angle cosine evolves from $\mu'$ to $\mu''$, and to order $\eta$,
\begin{align}
    {\mu''}^2& \simeq {\mu'}^2-(1-{\mu'}^2)\!\left(\frac{\Delta \updelta B_\parallel }{B_0}+\frac{1}{2}\frac{\Delta \updelta B_\perp^2}{B_0^2}\right)\!,
\label{eq:dmu3}
\end{align}
where $\Delta \updelta B_\parallel=\updelta B''_\parallel-\updelta B'_\parallel$ represents the change in the parallel component of the random contribution of the magnetic field, and $\Delta \updelta B_\perp=\updelta B''_\perp-\updelta B'_\perp$ represents the change in its perpendicular component. One important observation is that for slow modes, $\Delta \updelta B_\parallel/B_0\sim \!\left[\eta/(1-\eta)\right]^{1/2}\sqrt{2/3}$, while for Alfv\'en modes, $\Delta \updelta B_\parallel/B_0=\updelta B_\parallel=0$. For simplicity, we rewrite the second term in the rhs of Eq.~(\ref{eq:dmu3}) as $-(1-\mu^2)\delta$ in the following, with $\delta \sim \eta^{1/2}$ for slow modes and $\delta \sim \eta$ for Alfv\'en modes. To go further, it proves convenient to split the pitch-angle domain according to the sign, using here the symmetry $\mu\leftrightarrow-\mu$ of our estimates. We thus focus here on $\mu>0$. Moreover, we distinguish escaping particles with initial pitch angle cosine larger than $\delta^{1/2}$ and trapped ones.

The latter case corresponds to particles undergoing mirror reflections. The pitch-angle cosine of such particles becomes randomized with $\langle\mu\rangle\sim\delta^{1/2}$ and rms value $\langle\Delta\mu^2\rangle^{1/2}\sim\delta^{1/2}$, as indicated by Eq.~(\ref{eq:dmu3}). At values of $\delta$ not far below unity, this scaling holds for most particles.

For particles inside the loss cone, meaning $\mu_0\gtrsim\delta^{1/2}$, the pitch-angle cosine remains confined around the initial value, $\langle\mu\rangle\simeq\vert\mu_0\vert$, to within $\Delta\mu\simeq\delta/\vert\mu_0\vert$, since Eq.~(\ref{eq:dmu3}) then gives $\mu''\simeq\mu' - (1-{\mu'}^2)\Delta\updelta B_\parallel/(2\mu' B_0)$ to lowest order. 

As a consequence, the resonance is significantly narrower for particles with $\mu_0\gtrsim\delta^{1/2}$ than for those with $\mu_0\lesssim\delta^{1/2}$; hence, the pitch-angle averaged diffusion coefficients are strongly dominated by particles in the former range. We have checked this numerically as follows. We first performed a numerical study of a stochastic system that follows the evolution of the pitch-angle cosine through Eq.~(\ref{eq:dmu3}) step by step, accounting for mirroring whenever it occurs, in order to derive accurate estimates of $\langle\mu\rangle$ and $\langle\Delta\mu^2\rangle$. We then incorporated these estimates in the resonance function and computed the diffusion coefficients in various regimes of interest and compared the obtained values.

\begin{samepage}
In light of these studies, we find that the broadening of the resonance can be modeled, to the level of accuracy that suits the lowest-order expansion Eq.~(\ref{eq:prop1}), by assuming $\langle\mu\rangle \simeq \delta^{1/2}$ and $\langle\Delta\mu^2\rangle^{1/2}\sim\delta^{1/2}$. For comparison, Ref.~\cite{18Xu} uses $\langle\mu\rangle=\mu_0$ and $\langle\Delta\mu^2\rangle\sim\eta^{1/4}\sim\delta^{1/2}$ for all $\mu_0$, while Ref.~\cite{19Teraki} rather finds $\Delta\mu\sim\eta^{1/2}\sim\delta$. Which regime applies depends on $\eta$ and $\mu_0$, as explained above. Recall also that, for magnetosonic modes, $\delta^{1/2}\sim\eta^{1/4}$; hence, the amount of resonance broadening is significant, even in low-amplitude turbulence.

\section{NUMERICAL SIMULATIONS}\label{sec:num}
\subsection{Field prescription}
\label{sec:num_field}
The magnetic field is described as the sum of a static and uniform magnetic field  $\boldsymbol{B}_0$ and a turbulent component $\updelta \boldsymbol{B}$, expressed as the superposition of $N_{\bm k}$ MHD eigenmodes (pure fast, slow or Alfv\'en modes) with wave vector moduli between
$k_\text{min}$ and $k_\text{max}$. $k_\text{max}$ is chosen so that $k_\text{max}^{-1}<r_\text{g}$. Given that most of the energy of the turbulence is concentrated at the largest scales, the exact value of $k_\text{max}$ is not important and we typically take $k_\text{max}\sim 100\, r_\text{g}^{-1}$, while we set the mode density to $128$ per decade (which is rather conservative). 

\subsubsection{Isotropic fast mode turbulence}
\label{sec:num_iso}
For pure isotropic fast mode turbulence simulations, we draw $N_{\bm k}$ waves vectors with equally log-spaced norms and random directions and build the magnetic perturbations at the coordinates $(t,\boldsymbol{r})$ in the plasma rest frame as
\begin{equation}
    \updelta \boldsymbol{B}(t,\boldsymbol{r})=\sum_{i=1}^{N_{\bm k}} \updelta \boldsymbol{B}_{\boldsymbol{k}_i}^\text{F} \cos(\boldsymbol{k}_i\cdot\boldsymbol{r}-\omega_{i} t+\phi_{i}),
    \label{eq:ndB_F}
\end{equation}
where $\omega_i$ and $\updelta \boldsymbol{B}_{\boldsymbol{k}_i}^\text{F}$ are obtained by injecting the wave vector $\boldsymbol{k}_i$ in the dispersion and polarization relations~(\ref{eq:o_F}) and (\ref{eq:dB_F}), and $\phi_i$ is the random phase. The vectors are normalized according to the chosen spectral scaling and level of turbulence, namely for a Kolmogorov spectrum,
\begin{equation}
    \updelta b_{\boldsymbol{k}_i}^2=\frac{2\eta}{1-\eta}B_0^2 \left(\frac{k_i}{k_\text{min}}\right)^{-\frac{2}{3}} \left[\sum_{j=1}^{N_{\bm k}} \left(\frac{k_j}{k_\text{min}}\right)^{-\frac{2}{3}}\right]^{-1}\!.
\end{equation}
In the same fashion, we compute from Eq.~(\ref{eq:du_F}) the 4-velocity perturbation\footnote{For linear modes, there is no distinction between the 3-velocity and the spatial part of the 4-velocity but for finite amplitude modes, we treat $\delta \boldsymbol{u}$ as a 4-velocity to ensure that $|\delta \boldsymbol{E}|$ defined by Eq.~(\ref{eq:ndE_F}) remains smaller than $|\delta \boldsymbol{B}|$.} as 
\newpage
\end{samepage}

\begin{equation}
    \updelta \boldsymbol{u}(t,\boldsymbol{r})=\sum_{i=1}^{N_{\bm k}} \updelta \boldsymbol{u}_{\boldsymbol{k}_i}^\text{F} \cos(\boldsymbol{k}_i\cdot\boldsymbol{r}-\omega_{i} t+\phi_{i}),
\end{equation}
so that the electric field can be built according to the ideal MHD Ohm's law
\begin{equation}
    \updelta \boldsymbol{E}(t,\boldsymbol{r})=-\frac{\updelta\boldsymbol{u}(t,\boldsymbol{r})}{\sqrt{1+\updelta\boldsymbol{u}^2(t,\boldsymbol{r})}}\times \left[\boldsymbol{B}_0 +\delta \boldsymbol{B}(t,\boldsymbol{r})\right]\!.
    \label{eq:ndE_F}
\end{equation}

\subsubsection{Anisotropic slow and Alfv\'en mode turbulence}
\label{sec:num_GS}
For slow and Alfv\'en waves, we enforce \emph{approximately} the Goldreich-Sridhar scaling. The difficulty resides in that we need $\boldsymbol{k}_i$ to compute $\updelta \boldsymbol{B}_{\boldsymbol{k}_i}^\text{S}$ ($\updelta \boldsymbol{B}_{\boldsymbol{k}_i}^\text{A}$), but the direction of $\boldsymbol{k}_i$ depends on the direction of the \emph{local} magnetic field that we are trying to build, so a recursive procedure would be required. For simplicity, we assume that the direction of the local field can be approximated to that of $\boldsymbol{B_0}$. This approximation is reasonable if the perturbation is small enough at the scales of interest, that is, $\updelta B/B\lesssim k_\parallel/k\sim \left(k_\parallel/k_\text{min}\right)^{-1/2}$, which is increasingly constraining as we move to smaller scales. We therefore limit the dynamic range of these simulations and only consider $r_\text{g}k_\text{min}\gtrsim 10^{-2}$. For this limiting case and at scales close to the resonance condition $k_\parallel \sim r_\text{g}^{-1}$, we thus should have $\updelta B/B\lesssim0.1$ or $\eta\lesssim 10^{-2}$, which is the value that we adopt.

Depending on whether we consider undamped or damped waves, we use one of the following random field construction method.

For undamped waves, the procedure to construct the perturbed fields is essentially the same as the one described above for fast mode turbulence. $N_{\bm k}$ wave vectors are drawn with evenly log-spaced perpendicular components between $\sqrt{2/3}k_\text{min}$ and $k_\text{max}$ along random directions in the plane perpendicular to $B_0$ (in line with the remarks of the precedent paragraph), while the parallel components are defined as in Ref.~\citep{14Fatuzzo},
\begin{equation}
    k_{i\parallel}=\pm\frac{\sqrt{2}}{2}k_{i\perp}^{2/3}k_{\text{min}\perp}^{1/3}.
\end{equation}
For slow modes, we then take in combination with Eqs.~(\ref{eq:o_S}), (\ref{eq:dB_F}), and (\ref{eq:du_F}),
\begin{align}
    \updelta \boldsymbol{B}(t,\boldsymbol{r}) &=\sum_{i=1}^{N_{\bm k}} \updelta \boldsymbol{B}_{\boldsymbol{k}_i}^\text{S} \cos(\boldsymbol{k}_i\cdot\boldsymbol{r}-\omega_{i} t+\phi_{i}),\\
    \updelta \boldsymbol{u}(t,\boldsymbol{r}) &=\sum_{i=1}^{N_{\bm k}} \updelta \boldsymbol{u}_{\boldsymbol{k}_i}^\text{S} \cos(\boldsymbol{k}_i\cdot\boldsymbol{r}-\omega_{i} t+\phi_{i}),
\end{align}
with amplitudes scaling as
\begin{equation}
    \updelta b_{\boldsymbol{k}_i}^2=\frac{2\eta}{1-\eta}B_0^2 \left(\frac{k_{i\perp}}{k_{\text{min}\perp}}\right)^{-\frac{2}{3}} \left[\sum_{j=1}^{N_{\bm k}} \left(\frac{k_{j\perp}}{k_{\text{min}\perp}}\right)^{-\frac{2}{3}}\right]^{-1}\!,
\end{equation}
and construct the electric field as previously explained. A similar method is employed for Alfv\'en modes. 

For simulations with modes of finite time correlation, along the same lines as Ref.~\cite{16Hussein}, we add, compared to the previously described procedure, some pulsation spreading around the solution  $\omega_{\boldsymbol{k}_i}$ of the linear dispersion relation~(\ref{eq:o_S}),
\begin{align}
    \updelta \boldsymbol{B}(t,\boldsymbol{r}) &=\sum_{i=1}^{N_{\bm k}}\sum_{j=1}^{N_\omega} \updelta \boldsymbol{B}_{ij}^\text{S} \cos(\boldsymbol{k}_i\cdot\boldsymbol{r}-\omega_{ij} t+\phi_{ij}),\\
    \updelta \boldsymbol{u}(t,\boldsymbol{r}) &=\sum_{i=1}^{N_{\bm k}}\sum_{j=1}^{N_\omega} \updelta \boldsymbol{u}_{ij}^\text{S} \cos(\boldsymbol{k}_i\cdot\boldsymbol{r}-\omega_{ij} t+\phi_{ij}),
\end{align}
using the polarization given by Eqs.~(\ref{eq:dB_F}) and (\ref{eq:du_F}) for $\boldsymbol{k}_i$, the corresponding $\omega_{\boldsymbol{k}_i}$ and the magnitude
\begin{equation}
    \begin{split}
    \updelta b_{ij}^2={} &\frac{2\eta}{1-\eta}B_0^2 \Delta_i \mathcal{X}_{ij} \left(\frac{k_{i\perp}}{k_{\text{min}\perp}}\right)^{-\frac{2}{3}} \\
    &\times \left[\sum_{mn}\Delta_m \mathcal{X}_{mn} \left(\frac{k_{m\perp}}{k_{\text{min}\perp}}\right)^{-\frac{2}{3}}\right]^{-1},
    \end{split}
\end{equation}
where $\Delta_i$ is the constant logarithmic spacing of $\omega_{ij}$ at fixed $i$ and $\mathcal{X}_{ij}$ is the function describing the decorrelation of modes away from $\omega_{\boldsymbol{k}_i}$ and whose exact form is model dependent. 
To make contact with our theoretical calculations, we choose
\begin{equation}
    \mathcal{X}_{ij}=\frac{\gamma_\text{d} |\omega_{\boldsymbol{k}_i}|}{\uppi} \frac{|\omega_{ij}|}{\gamma_\text{d}^2 \omega_{\boldsymbol{k}_i}^2+\left(\omega_{ij}-\omega_{\boldsymbol{k}_i}\right)^2},
\end{equation}
which is the Fourier transform of $t\mapsto \e^{\i\omega_{\boldsymbol{k}_i}t-\gamma_\text{d}|\omega_{\boldsymbol{k}_i}|t}$, and corresponds to the "Nonlinear Anisotropic Dynamical Turbulence" model of \cite{16Hussein}. The pulsation cutoff is chosen to be much larger than the linear pulsation (at least $1000\, |\omega_{\boldsymbol{k}_i}|$) and the mode density to at least eight modes per decade. 

\subsection{Measurements}
\label{sec:meas}
For a given physical setup and particle rigidity, a thousand of ultrarelativistic particles are tracked, each injected with a random initial pitch angle in a different turbulence realization. The Bulirsch-Stoer integrator \citep{96Press} is used to evolve the positions $\boldsymbol{r}$ and 4-velocities $\Gamma\boldsymbol{\beta}$ in the plasma rest frame, according to the equations of motion 
\begin{equation}
\begin{aligned}
   &\frac{\mathrm{d}\boldsymbol{r}}{\mathrm{d}t}=\boldsymbol{\beta},\\ 
  &\frac{\mathrm{d}\Gamma\boldsymbol{\beta}}{\mathrm{d}t}=\frac{q}{m}\left(\updelta\boldsymbol{E} + \boldsymbol{\beta}\times\boldsymbol{B}\right),
\end{aligned} 
\label{eq:motion}
\end{equation}
where $q$ and $m$ are the electric charge and the rest mass of the particle and $\boldsymbol{B}=\boldsymbol{B}_0+\updelta \boldsymbol{B}$ is the total magnetic field, constructed according to the procedure presented above.

For most simulations, the initial momentum distribution of the particles is a Dirac-delta function in the simulation frame so that all particles have the same initial $r_\text{g}k_\text{min}$ in that frame. For fast mode turbulence simulations with $\beta_\text{F}\geq 0.7 $ (and $\eta=0.3$), we also carried out simulations such that the gyroradius distribution is a delta function in the local rest frame of the fluid (in which the motional electric field is initially zero) in order to be closer to the situation described by our analytical calculations. If $r'_\text{g,0}=p'_0/(qB')$ is the value of the gyroradius to be probed in that frame, then in the simulation frame, we initialize the particles with a momentum
\begin{equation}
    p_0=\frac{p'_0}{\gamma_u(1-\beta_u \mu_u)}=\frac{q\,(B_\perp^2/\gamma_u^2+B_\parallel^2)^{1/2}}{\gamma_u(1-\beta_u \mu_u)}r'_\text{g,0},
\end{equation}
where $\beta_u$ ($\gamma_u$) denotes the velocity (the Lorentz factor) of the fluid at the initial position of the particle, $\mu_u$ is the cosine of the angle its initial momentum makes with the flow and $B_\perp$ ($B_\parallel$) is the perpendicular (parallel) component of the local magnetic field. For these simulations, we report the measurements of the diffusion coefficients (Figs.~\ref{plot:F_Dmu} and~\ref{plot:F_Dp}) at an effective gyroradius that is defined as the mean momentum, after any transient initial energy kick, divided by $qB_0$.

\subsubsection{Pitch-angle diffusion coefficient}
The pitch-angle diffusion coefficient is estimated by computing the mean square pitch-angle displacement $\langle[\Delta \mu(t)]^2\rangle/2t$, where $\langle \ldots \rangle$ denotes the average over the particles, i.e., over the pitch angles and the turbulence realizations. We fit $\langle[\Delta \mu(t)]^2\rangle$ using least squares to a linear function at early times (but still larger than $r_\text{g}$ to omit the ballistic regime) and identify $D_{\mu\mu}$ to one-half of the slope of the fitting function. One limitation is that irrespective of the energy dependence of the diffusion coefficients, $\langle[\Delta \mu(t)]^2\rangle/2t$ eventually displays a subdiffusive behaviour as $t$ increases due to the bounded nature of $\Delta \mu(t)$ and a well-defined plateau cannot always be observed. 

\begin{figure}
\center \includegraphics[scale=1.]{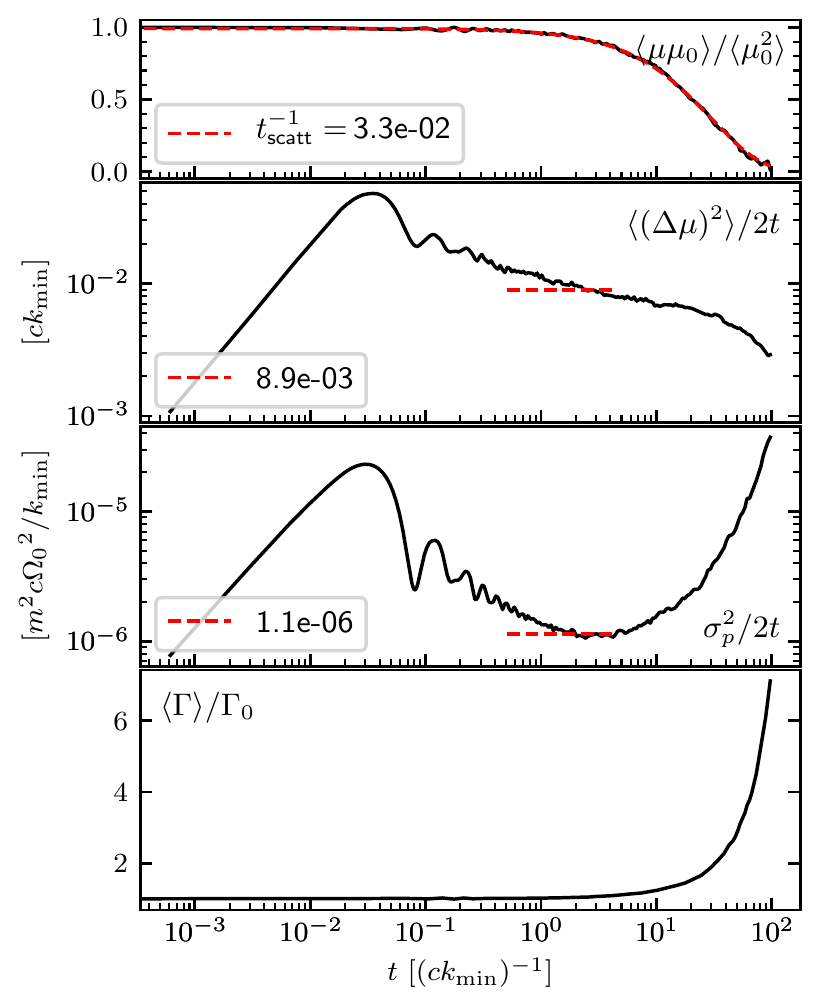}
\caption{From top to bottom, pitch-angle correlation function; pitch-angle mean square displacement; standard deviation of the distribution of $p$, ratio of the mean Lorentz factor to its initial value, as a function of time, for the simulation of damped Alfv\'en modes at the smallest rigidity. The red dashed lines indicates the results of the fits discussed in the text.}
\label{plot:meas-eg}
\end{figure}

These results were cross-checked against another estimation relying on the pitch-angle correlation function $\langle\mu(t) \mu(0) \rangle$, where the data are fitted against a decaying exponential model $\propto \exp(-t/t_\text{scatt})$ and $D_{\mu\mu}$ is estimated as $1/t_\text{scatt}$. Most results were found to differ by a factor of a few\footnote{In details, when derived from the correlation function, $D_{\mu\mu}$ is found to be $\approx 3$ times larger for fast modes,  $\approx 4$ times larger for Alfv\'en modes, $\approx$ equal for damped slow modes, and $\approx 1.7$ times smaller for undamped slow modes.} (see, for instance, Fig.~\ref{plot:meas-eg}), although for a few simulations (namely that of fast modes with $\beta_F\gtrsim 0.7$, $r_\text{g}k_\text{min}\gtrsim 1$), the data are not well fitted by an exponential model. We also note that this method requires longer computation times than when using the running displacement. 

\subsubsection{Momentum diffusion coefficient and acceleration timescale}
\begin{figure}
\center \includegraphics[scale=1.]{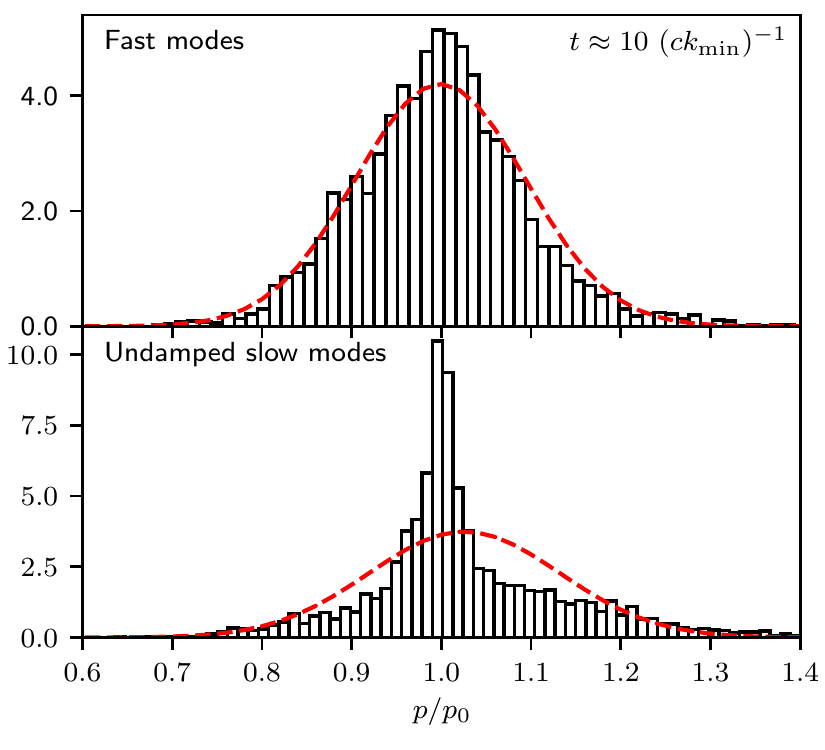}
\caption{Histograms of $p$ at $t\approx 10^3 r_\text{g}$ for particles injected in isotropic fast mode turbulence (top) and undamped slow mode turbulence (bottom). The data at times within $\pm 1\%$ of each other were used to increase the number of events (5120). The red dashed lines depict the normal distributions corresponding to the mean and variance values of the data. }
\label{plot:his-eg}
\end{figure}

In a similar fashion, the pitch-angle averaged momentum diffusion coefficient is evaluated from the dispersion of the energy distribution with $D_{pp}$ set to the slope of the linear fit of $\sigma_p^2(t)/2$ where $\sigma_p$ is the standard deviation of the distribution of $p$ at early times. For subrelativistic simulations, this is equivalent to using the mean square displacement around the initial value $\langle[\Delta p(t)]^2\rangle/2t$. For relativistic setups, particles are subject to a quick initial increase of energy due to some first order Fermi acceleration; hence, using $\langle[\Delta p(t)]^2\rangle/2t$ would lead to overestimating $D_{pp}$. This initial acceleration moreover implies that we do not measure the diffusion coefficient for the initial $r_\text{g}k_\text{min}$ but rather a slightly larger value (approximately, $1.4$ larger for $\beta_\text{A}=0.7$ and $1.6$ larger for $\beta_\text{A}=0.9$, when the particle momentum distribution is initialized in the simulation frame).
Finally, we note that the energy distribution is not always well described by a Gaussian, e.g., Refs.~\cite{14Fatuzzo,19Teraki}. In particular, for undamped slow modes turbulence in which there are two different populations of particles (those sensitive to resonance broadening and those which are not), the distributions display large wings (see, for instance, Fig.~\ref{plot:his-eg}), and the above described procedure leads to overestimating the acceleration efficiency.

In light of this last remark, the pitch-angle cosine averaged acceleration timescale is not derived from $D_{pp}$ but defined as the time when the average Lorentz factor has doubled, $\langle\Gamma (t_\text{acc}) \rangle=2 \Gamma_0$. Both approaches lead to similar results for most simulations anyway\footnote{In details, for $r_\text{g}k_\text{min}<1$, $2t_\text{acc}D_{pp}/p^2$ takes the following values. For fast modes, $\approx 0.4$ at $\beta_\text{A}=10^{-2}$, $\approx 0.5$ at $\beta_\text{A}=10^{-1}$, and $\approx 2$ at $\beta_\text{A}\geq 0.7$.  For undamped and damped Alfv\'en modes, $\approx 0.7$ and $\approx 0.5$, respectively. For undamped and damped slow modes, $\approx 4$ and $\approx 0.4$, respectively. For $r_\text{g}k_\text{min}\approx 1$, $1 \lesssim t_\text{acc}D_{pp}/2p^2 \lesssim 1.5$, regardless of the turbulence type.}.

\section{POWER SPECTRUM OF ELECTRIC FIELD FLUCTUATIONS\\ IN WAVE TURBULENCE}\label{sec:appE}
In a pair plasma, inertia effects appear in the Ohm's law and generate the following extra electric field component~\cite{1996PhRvL..76.3340G},
\begin{equation}
\updelta\boldsymbol{E^\times} \simeq \frac{\partial}{\partial t}\left(\frac{w}{2e^2n^2}\updelta\boldsymbol{ j}\right)\!,
\label{eq:Ohm1}
\end{equation}
emphasizing that $\updelta\boldsymbol{ E^\times}$ is here measured in the plasma rest frame; $w$ represents the enthalpy density, $n$ is the number density, and $\updelta\boldsymbol{ j}$ is the total current density. In Fourier space, relating $\updelta\boldsymbol{ j}$ to $\updelta\boldsymbol{ B}$ to lowest order in $\omega/k$, we obtain for this extra electric field component,
\begin{equation}
\updelta\boldsymbol{E^\times_k} = \kappa \frac{\omega}{4\uppi\omega_\text{p}^2}\,\boldsymbol{k}\times\updelta\boldsymbol{B_k},
\label{eq:Ohm2}
\end{equation}
where $\kappa = \left[1+\frac{T}{m}\frac{\hat\Gamma}{\hat\Gamma-1}\right]$,  $\hat\Gamma$ denotes the adiabatic index and, $T$ is the temperature. The scaling $\updelta E^\times \propto \omega k \updelta B_k$ confirms that the parallel electric field power is maximum on the smallest length scales of the turbulent cascade.

Although each component of this small-scale electric field is orthogonal to the corresponding magnetic field component in Fourier space, this is no longer the case once the sum over $\boldsymbol{k}$ is performed. More specifically, $\left\langle \updelta\boldsymbol{E^\times}\cdot\boldsymbol{B_0}\right\rangle = 0$ and $\left\langle \updelta\boldsymbol{E^\times}\cdot\updelta\boldsymbol{B}\right\rangle = 0$ for all modes, but $\left\langle \left(\updelta\boldsymbol{E^\times} \cdot\boldsymbol{B_0}\right)^2\right\rangle \neq 0$ and $\left\langle \left(\updelta\boldsymbol{E^\times}\cdot\updelta\boldsymbol{ B}\right)^2\right\rangle \neq 0$ for Alfv\'en polarization, while $\left\langle \left(\updelta\boldsymbol{E^\times} \cdot\boldsymbol{B_0}\right)^2\right\rangle = 0$ but $\left\langle \left(\updelta\boldsymbol{E^\times}\cdot\updelta\boldsymbol{ B}\right)^2\right\rangle \neq 0$ for magnetosonic modes.

Using the polarization of each mode and the corresponding spectrum described earlier, one can calculate the rms parallel electric field $\updelta E_\parallel = \left\langle (\updelta\boldsymbol{ E^\times} \cdot\boldsymbol{B})^2\right\rangle^{1/2}\!\big/B$, recalling that $\boldsymbol{B} = \boldsymbol{B_0}+\updelta\boldsymbol{B}$. 

For an isotropic cascade of fast modes, we derive
\begin{equation}
\updelta E^\text{F}_\parallel \simeq
\frac{\kappa}{4\uppi\sqrt{30}}\,\beta_\text{F}\,\eta\,\frac{k_\text{max}^{5/3}k_\text{min}^{1/3}}{\omega_\text{p}^2}\,B.
\label{eq:EparF}
\end{equation}

For a Goldreich-Sridhar cascade of slow modes,
\begin{equation}
\updelta E^\text{S}_\parallel \simeq
\frac{\kappa}{16\uppi}\,\beta_\text{S}\,\eta\,\frac{k_\text{max}^{4/3}k_\text{min}^{2/3}}{\omega_\text{p}^2}\,B.
\label{eq:EparS}
\end{equation}

Finally, for a Goldreich-Sridhar cascade of Alfv\'en modes, we obtain
\begin{align}
\updelta E^\text{A}_\parallel \simeq &
\frac{\kappa}{8\uppi}\,\beta_\text{A}\,\frac{k_\text{max}k_\text{min}}{\omega_\text{p}^2}\nonumber\\
&\times B\,\biggl[\sqrt{\frac{2}{3}}\eta+\sqrt{\eta(1-\eta)}\left(\frac{k_\text{max}}{k_\text{min}}\right)^{1/3}
\biggr].
\label{eq:EparA}
\end{align}
The second term in brackets corresponds to the component parallel to $B_0$. 

In the case of electron-ion plasmas, the above corrections are completed by the Biermann battery term, which provides the additional electric field
\begin{equation}
\updelta\boldsymbol{E^\times_k}=\i\boldsymbol{k}\frac{{\updelta p_e}_k}{n}
\label{eq:EparBB1}
\end{equation}
in terms of the electron pressure fluctuation ${\updelta p_e}_k$. Using ${\updelta p_e}_k \sim \hat\Gamma\frac{T_e}{m_p}\boldsymbol{k}\cdot\updelta\boldsymbol{\beta_k}/\omega$, as applies to linear magnetosonic modes, we find here as well that the power of $\updelta E_\parallel$ peaks at small scales. Its contribution is dominated by the product $\updelta\boldsymbol{E^\times_k}\cdot\boldsymbol{B_0}$, which gives
\begin{equation}
\updelta E^\text{F}_\parallel \simeq \frac{\hat\Gamma}{\sqrt{6}}\frac{T_e}{m_p}\,\frac{k_\text{max}^{2/3}k_\text{min}^{1/3}}{\omega_\text{ci}}\,\sqrt{\eta(1-\eta)}\,B
\label{eq:EparBB2}
\end{equation}
for fast modes and 
\begin{align}
\updelta E^\text{S}_\parallel \simeq & \frac{\hat\Gamma\,\beta_\text{S}^2}{2\sqrt{2}\beta_\text{s}^2}\frac{T_e}{m_p}\,\frac{k_\text{max}^{1/3}k_\text{min}^{2/3}}{\omega_\text{ci}}\,\sqrt{\eta(1-\eta)}\,B\nonumber\\
&\times\left[2\sqrt{2}+\sqrt{\frac{\eta}{1-\eta}}\left(\frac{k_\text{max}}{k_\text{min}}\right)^{1/3}\right]
\label{eq:EparBB3}
\end{align}
for slow modes, in terms of $\omega_\text{ci} = q B/(m_p c)$, the proton cyclotron frequency.

The various dependencies on $k_\text{max}$ of the above estimates provide the scale dependence of the corresponding electric field; i.e., if the spectral index of the (one-dimensional) power spectrum of $\updelta E_\parallel$ is $q_E$, meaning $k^2 \langle\vert{\updelta E_\parallel}_k\vert^2\rangle \propto k^{-q_E}$, then $\updelta E_\parallel \propto  k_\text{max}^{(1-q_E)/2}$, assuming $q_E<1$. In detail, Eq.~(\ref{eq:EparF}) for fast modes leads to $q_E=-7/3$, 
Eq.~(\ref{eq:EparS}) for slow modes leads to $q_E=-5/3$, and
Eq.~(\ref{eq:EparA}) for Alfv\'en modes leads to $q_E=-1$. Equivalently, the parallel electric field rms strength on a given scale $k$ scales as $\left.\updelta E_\parallel\right\vert_k \propto k^{(1-q_E)/2}$, to be contrasted with a typical $\left.\updelta E\right\vert_k \propto k^{-1/3}$ for the MHD component. Consequently, the parallel electric field energy may represent a non-negligible contribution to the total electromagnetic energy density, but in all cases studied above, its power is strongly suppressed on the largest scales.

\bibliography{refs}

\end{document}